\documentclass[aps,prl,twocolumn,nopacs,superscriptaddress,longbibliography]{revtex4-1}
\usepackage{color}
\usepackage{graphicx}
\usepackage{mathrsfs}
\usepackage{bm}
\usepackage{lipsum}
\usepackage{sidecap}
\sidecaptionvpos{figure}{t}
\usepackage[english]{babel}
\usepackage[dvipsnames]{xcolor}
\usepackage{xfrac}

\usepackage[%
colorlinks=true,
urlcolor=RoyalBlue,
linkcolor=RoyalBlue,
citecolor=RoyalBlue,
]{hyperref}

\usepackage[type1]{libertine}                                        
\usepackage{textcomp}
\usepackage[scaled=.85]{beramono}
\usepackage[libertine,cmintegrals,cmbraces,vvarbb,slantedGreek]{newtxmath}
\usepackage[scr=boondoxo]{mathalfa}
\usepackage{bm}
\usepackage[lf]{carlito}

\newcommand{\figref}[1]{Fig.\,\ref{#1}}
\newcommand{\spinhalf}{spin-\sfrac{1}{2}}

\begin{document}

\title{Combining Electron Spin Resonance Spectroscopy with Scanning Tunneling Microscopy at High Magnetic Fields}

\author{Robert Drost}
\author{Maximilian Uhl}
\author{Piotr Kot}
\author{Janis Siebrecht}
\affiliation{Max-Planck-Institute for Solid State Research, Heisenbergstr. 1, 70569 Stuttgart, Germany}

\author{Alexander Schmid}
\author{Jonas Merkt}
\author{Stefan W\"unsch}
\author{Michael Siegel}
\affiliation{Institut f\"ur Mikro- und Nanoelektronische Systeme, Karlsruhe Institute of Technology, Hertzstr. 16, 76187 Karlsruhe, Germany}

\author{Oliver Kieler}
\affiliation{Physikalisch-Technische Bundesanstalt, Bundesallee 100, 38116 Braunschweig, Germany}

\author{Reinhold Kleiner}
\affiliation{Physikalisches Institut, Center for Quantum Science (CQ) and LISA$^+$, Universität Tübingen, 72076 Tübingen, Germany}

\author{Christian R.\ Ast}
\email{c.ast@fkf.mpg.de}
\affiliation{Max-Planck-Institute for Solid State Research, Heisenbergstr. 1, 70569 Stuttgart, Germany}

\begin{abstract}

Magnetic media remain a key in information storage and processing. The continuous increase of storage densities and the desire for quantum memories and computers pushes the limits of magnetic characterisation techniques. Ultimately, a tool which is capable of coherently manipulating and detecting individual quantum spins is needed. The scanning tunnelling microscope (STM) is the only technique which unites the prerequisites of high spatial and energy resolution, low temperature and high magnetic fields to achieve this goal. Limitations in the available frequency range for electron spin resonance STM (ESR-STM) mean that many instruments operate in the thermal noise regime. We resolve challenges in signal delivery to extend the operational frequency range of ESR-STM by more than a factor of two and up to 100\,GHz, making the Zeeman energy the dominant energy scale at achievable cryogenic temperatures of a few hundred millikelvin. We present a general method for augmenting existing instruments into ESR-STMs to investigate spin dynamics in the high-field limit. We demonstrate the performance of the instrument by analysing inelastic tunnelling in a junction driven by a microwave signal and provide proof of principle measurements for ESR-STM.

\end{abstract}

\maketitle

\section{Introduction}

The direct manipulation and detection of individual spins (see \figref{fig:Figure0}(a)) is one of the major goals in contemporary nanoscience \cite{heinrich_single-atom_2004, hirjibehedin_large_2007, baumann_electron_2015, willke_probing_2018, yang_coherent_2019, seifert_single-atom_2020, veldman_free_2021,balatsky_electron_2012,mullegger_radio_2014}. Meeting these challenges requires a local measurement of electronic and magnetic properties with atomic precision. The scanning tunnelling microscope (STM) routinely achieves this limit of resolution and is thus an ideal tool to study the dynamics of magnetic nano-objects \cite{binnig_surface_1982} on their own length and time scales \cite{loth_measurement_2010, cocker_ultrafast_2013, yoshida_probing_2014,gutzler_light-matter_2021}.

Combining electron spin resonance with STM (ESR-STM) has introduced new possibilities to the local studies of individual spins and has expanded the available parameters space substantially, but it imposes a series of strict experimental requirements, most notably on the base temperature of the cryostat. The operational frequency range of the instrument determines the maximum magnetic field for ESR-STM experiments and sets the relevant energy scale in the experiment. ESR-STM relies on the thermal initialisation of the target systems into their ground state. However, in many contemporary implementations of ESR-STM, the Zeeman energy is on the order of $k_B T$ and a non-negligible excited state population remains \cite{willke_probing_2018,seifert_single-atom_2020}. This is a significant impediment to resolving intrinsic spin dynamics at the nanoscale. The goal of coherent manipulation from a known ground state may be reached via two approaches: Reducing the base temperature of the experiment to suppress thermal excitations from the ground state, or increase the microwave frequency to operate at higher magnetic fields.
 
Current implementations of ESR-STM typically operate at frequencies up to 40\,GHz \cite{paul_generation_2016,natterer_upgrade_2019,friedlein_radio-frequency_2019}. To achieve thermal initialisation of the target systems at these frequencies mK temperatures are required, which are only achievable in dilution refrigerators \cite{weerdenburg_scanning_2021}. This approach requires dedicated machines that are costly to produce and present significant challenges in everyday operation. High frequency signals in the upper GHz range, on the other hand, can be generated in an independent setup outside the ultra-high vacuum (UHV) system and routed to the tunnel junction through a set of suitable cables. This approach is, therefore, more flexible, allowing the retrofitting or modification of existing machines by the addition of dedicated high GHz cabling \cite{kot_microwave-assisted_2020}.

Extending the operational frequency range of ESR-STM has thus far been prevented by challenges in signal delivery. We have augmented a commercially available STM (Unisoku model USM1300) featuring 310\,mK base temperature and a 6\,T single axis magnet with an antenna assembly which permits us to deliver microwave signals of up to 105\,GHz directly to the tunnel junction. This work can be used as a guideline to design new instruments or retrofit existing ones for high GHz microwave capabilities.

\section{Instrument Design}

The Unisoku model USM1300 STM is a commercially available experimental platform combining ultrahigh vacuum (UHV) sample preparation and ultra-low temperature STM with high field capabilities. The STM unit, developed and manufactured by the Unisoku corporation, is installed on the insert of a superinsulated $^4$He bath cryostat produced by Cryogenic Ltd. The insert includes a 1\,K pot, which is supplied with liquid helium from the bath through an adjustable needle valve, and a single-shot $^3$He cooling cycle. With a total volume of 30\,l $^3$He gas, the STM is capable of operating at a base temperature of 310\,mK for up to 72 hours.

\begin{figure}
	\centering
		\includegraphics[width=0.8\columnwidth]{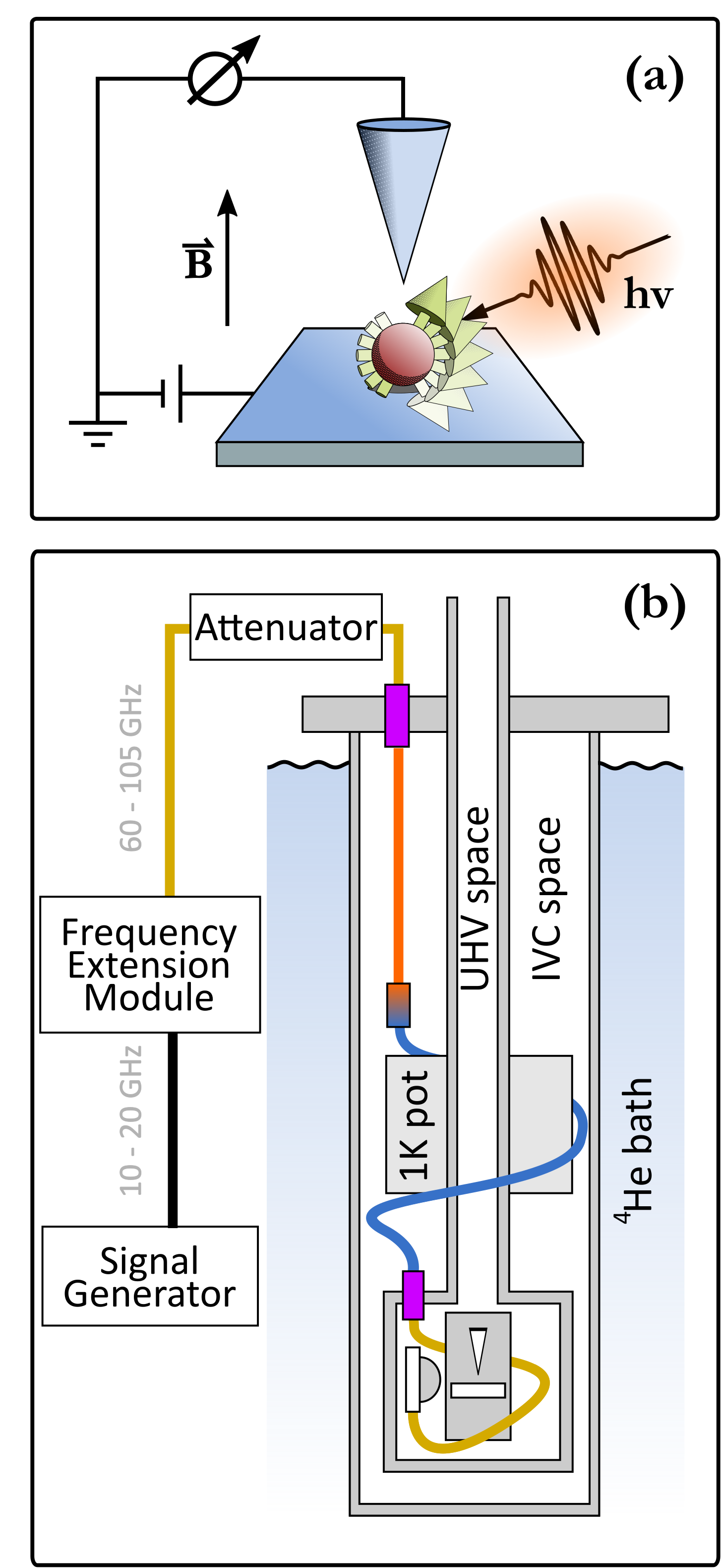}
		\caption{(a) Principle of an ESR-STM measurement: An external microwave signal induces spin precession in a magnetic atom probed by the STM tip. (b) Sketch of the cryostat insert housing the STM with the RF wiring and antenna we installed in the system. Black lines indicate SMA cables, yellow lines flexible .047 cables, blue lines Cu/SPCW .047 cables, and orange lines superconducting NbTi/NbTi .047 cables. Feedthroughs are indicated in purple. The $^3$He stage is omitted  for clarity.}
	\label{fig:Figure0}
\end{figure}

Figure \ref{fig:Figure0}(b) shows a sketch of the cryostat insert including the modifications we implemented as part of the ESR-STM augmentation. The main addition to the base setup is the installation of a series of 0.047\,inch semirigid coaxial cables, rated to 110\,GHz, and a radio frequency antenna into the system. We solve the challenges of finding leak-tight vacuum feedthroughs and thermalisation of the RF assembly to produce a high-performance machine capable of delivering high-frequency signals at large amplitudes onto the tunnel junction. Our approach extends the operational frequency range of ESR-STM by a factor of more than two while maintaining signal amplitudes comparable with previous efforts. Below is a step-by-step discussion of the design philosophy and implementation of our custom modifications to the base system.

\begin{figure}
	\centering
		\includegraphics[width=1.00\columnwidth]{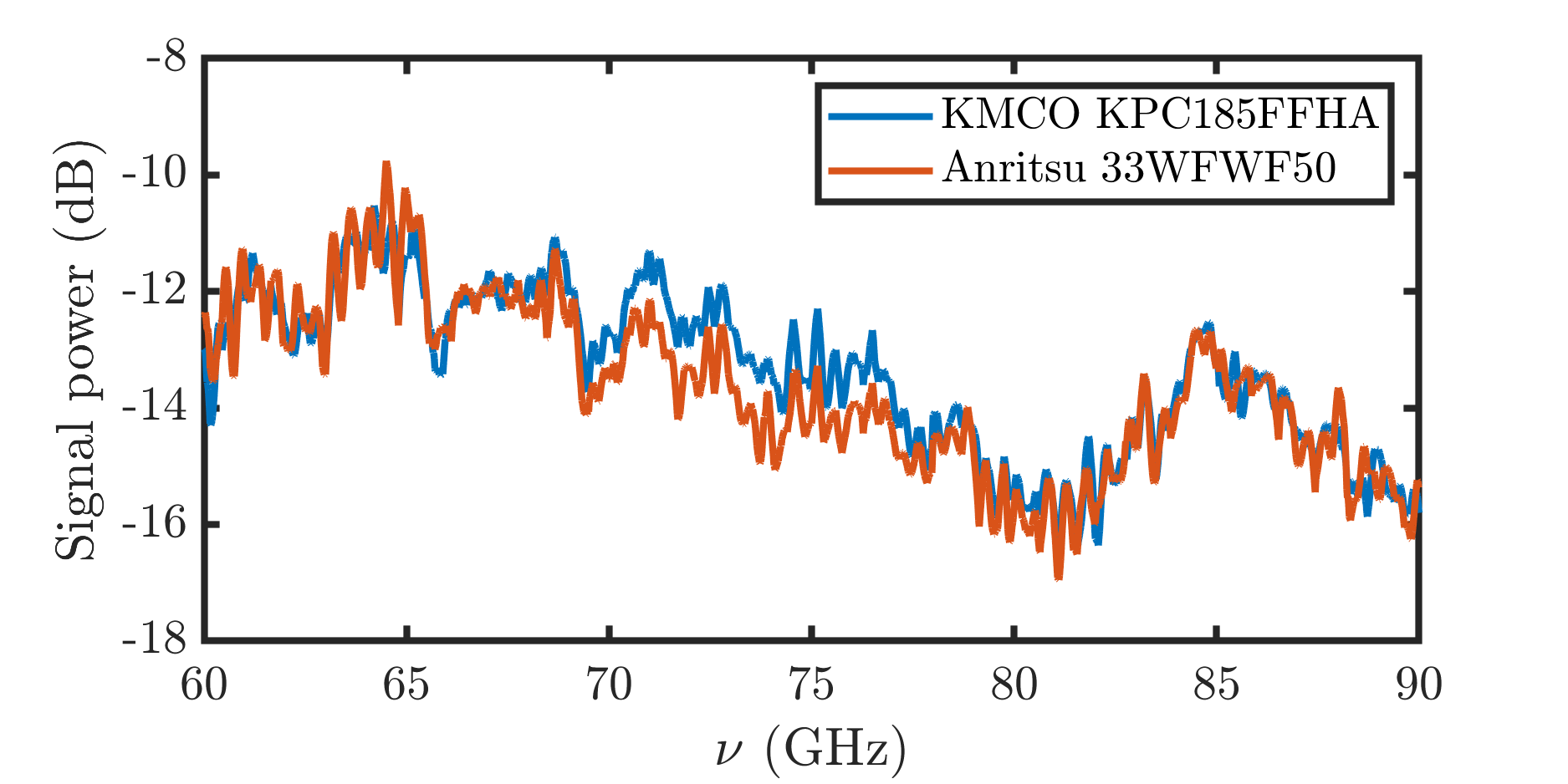}
		\caption{Transmission through the 1.85\,mm feedthrough KPC185FFHA, rated to 65\,GHz, (blue) and the 1\,mm female-female adapter 33WFWF50, rated to 110\,GHz. Bot components have comparable losses in the frequency range of interest.}
	\label{fig:Figure1}
\end{figure}

\subsection{High-frequency wiring}

The geometry of STMs presents a severe challenge to the integration of high-frequency signals. The confined space of the scanner housing and complex shape of the tip-sample system thwart any attempt to realise an impedance-matched connection to the tip. The resulting scattering of electromagnetic waves in the instrument will inevitably lead to high losses. This makes it all the more difficult to bring a high amplitude signal close to the tunnel contact. The requirement for high signal amplitudes leads to a conflict with a key design principle for cryostats to use high-resistance cabling in order to limit the thermal load on the experiment.

We overcome these issues by using a combination of high conductance coaxial cables from different materials to achieve maximum power transmission to the antenna. We use semirigid cables with a copper jacket and silver-plated copper weld (SPCW) conductor on the air side and in the upper sections of the cryostat to the 1K-pot (see red wire section in \figref{fig:Figure0}(b)). Their high conductance ensures small signal losses even at room temperature. We installed a coaxial cable with NbTi shield and conductor running from the 1K-pot to the $^3$He stage (see blue wire section in \figref{fig:Figure0}(b)). NbTi is a superconductor with a transition temperature of 10\,K and a high critical field of 15\,T. As superconductors are excellent conductors of electricity, but very poor conductors of heat \cite{tinkham_superconductivity_1994}, the NbTi cable provides excellent signal transmission at low temperatures while essentially eliminating thermal loads on the low-temperature parts of the experiment. Finally, a flexible coaxial cable with silver plated copper shield and conductor connects the RF antenna in order to preserve STM motion during sample transfer and spring damping during regular operation.

We use semirigid coaxial cables with 0.047\,in outer diameter for all applications. This cabling standard is rated to 110\,GHz with an impedance of 50\,$\upOmega$. We installed 1\,mm connectors (Anritsu W1 series), also rated to 110\,GHz, on all cable segments. Before installation in the machine, all cables were repeatedly immersed in liquid helium and rigorously tested for any temperature related damages. 

Our combination of copper/SPCW and superconducting cables requires excellent heat management in order to work effectively. To ensure proper thermalisation, all cables are anchored at several points inside the cryotat. The thermal anchors consist of a copper braid, glued to the outer conductor of the cable over a large surface area using thermally conductive silver epoxy, and fastened to the anchor points with a screw and lug. The copper/SPCW cable is anchored at the baffles of the cryostat, the sorption pump, and the 1K-pot. The NbTi cable is thermally anchored at the 1K-pot and $^3$He-pot as described above and wound in a wide loop around the UHV column to accommodate the thermal expansion and contraction of the wiring during cool-down or warm-up. The installation of our custom radio frequency cabling did not affect the base temperature of the instrument.

\subsection{High-frequency UHV feedthrough}

To our knowledge, there are no commercially available hermetically sealed double ended coaxial vacuum feedthroughs rated to 90\,GHz or above. In practice, however, vacuum feedthroughs with 1.85\,mm connectors, rated to 65\,GHz, show very low losses up to at least 90\,GHz and can act as a substitute. \figref{fig:Figure1} shows transmission measurements through the hermetically sealed feedthrough KPC185FFHA by Kawashima Manufacturing Corporation from 60\,GHz to 90\,GHz. The performance of KPC185FFHA is comparable to 1\,mm female-femals adapters rated to 110\,GHz.  A pair of 1.85\,mm to 1\,mm adapters (e.g. CentricRF C8186) is needed to mate the feedthrough to the high-frequency cabling. 

\subsection{High-frequency antenna}

The implementation of the dedicated high-frequency line requires a solution for coupling the high-frequency signal to the tunnel junction. We designed an antenna to transform the electric signal in the high-frequency line into electro-magnetic radiation illuminating the tunnel junction, with the STM tip effectively acting as a receiver. A dedicated antenna ensures efficient coupling into the vacuum, eliminating losses. Mounting it in close proximity to the tunnel contact further improves the signal strength.

\begin{figure}
	\centering
		\includegraphics[width=1.00\columnwidth]{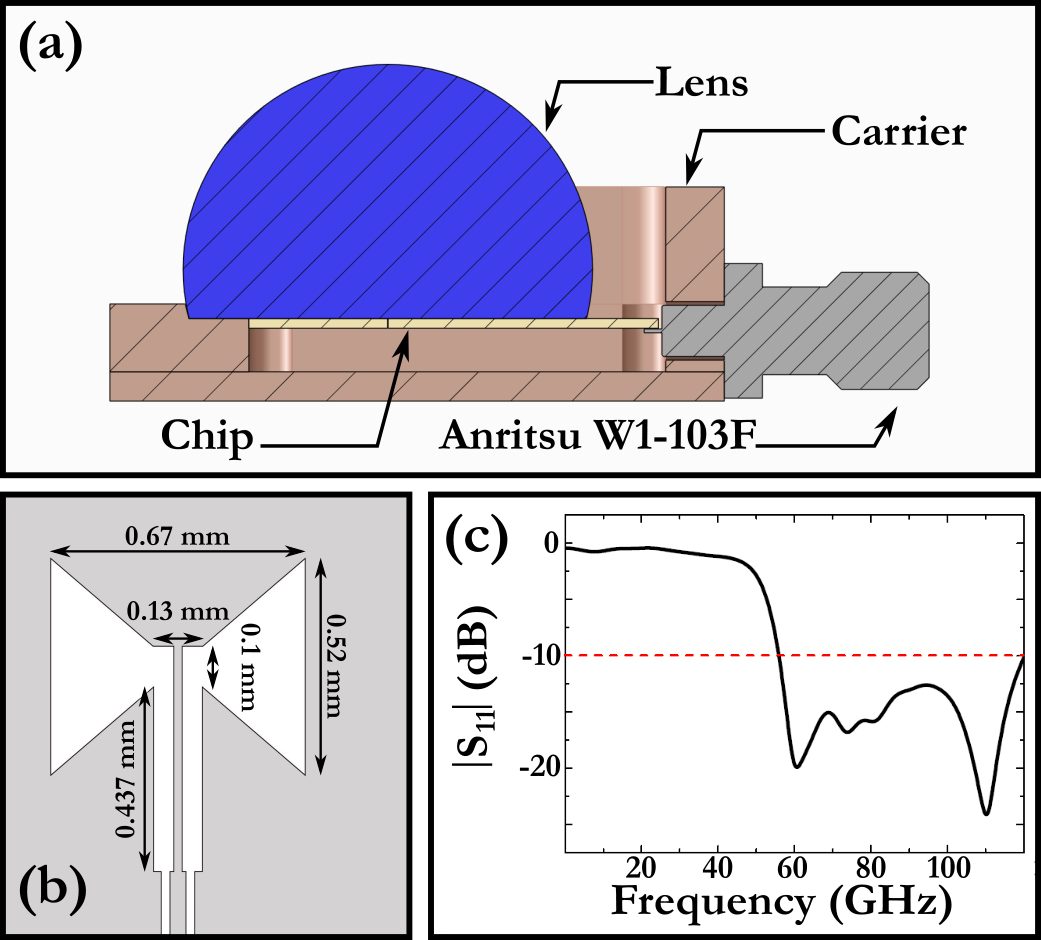}
	\caption{Antenna design: \textbf{(a)} Sectional drawing of the antenna carrier with the lens installed. \textbf{(b)} Scale drawing of the antenna. Grey areas denote the AuPd layer. The antenna is assembled on a 10$\times$20\,mm$^2$ high resistivity silicon chip. \textbf{(c)} Simulated reflectance of the antenna up to 120\,GHz. The antenna achieves excellent power dissipation across a wide band beginning at about 60\,GHz.}
	\label{fig:Figure3}
\end{figure}

We chose an on-chip antenna design for a compact and integrated solution \cite{merkt_entwurf_2017}. The silicon chip is mounted in a phosphorous bronze carrier attached directly to the side of the STM scanner housing, shown in \figref{fig:Figure3}(a). A flange mount connector (Anritsu W1-103F) provides the electrical contact to the high-frequency wiring. To increase the microwave power incident on the tunnel junction, we fitted the carrier with a hyper-hemispherical silicon lens flush with the underside of the antenna chip to partially collimate the radiation.

A broadband bowtie antenna affords the most flexibility, covering the entire intended frequency range of 60 to 90\,GHz. \figref{fig:Figure3}(b) shows a dimensional drawing of the optimised antenna structure installed in the microscope. The antenna is assembled in a thin AuPd film on a high resistivity silicon substrate of 380\,$\upmu$m thickness. Design and parameter optimisation for the antenna was performed using the CST Microwave Studio software with performance tests in 8.4:1 scale models. \figref{fig:Figure3}(c) shows the simulated reflectance of the antenna from DC up to 120\,GHz \cite{merkt_entwurf_2017}. The design achieves excellent power dissipation across a wide band beginning at about 60\,GHz.

\section{Radio Frequency Generation}

We use a multi-stage generation scheme to reach our intended operational frequency window (see \figref{fig:Figure0}(b)). The first stage is a baseband generator (Keysight 8257D) capable of producing signals up to 20\,GHz. This generator feeds a frequency extension module (VDI WR12SGX), which multiplies the input frequency by a factor of six. The frequency extension module is intended to operate with input frequencies between 10 and 15\,GHz, producing a constant amplitude signal between 60 and 90\,GHz. We found that upper limit can be extended by feeding the module with higher input frequencies, but at a significant cost in amplitude. Still, overdriving allows us to extend the operational frequency range of our ESR-STM to 105\,GHz. Expanding the frequency range at the lower end in the same fashion is not possible due to the sharp cut-off of the WR12 waveguide output on the extension module. We regulate the source amplitude through a computer-controlled rotary vane attenuator (Mi-Wave 511E/387ND). This device allows us to regulate the power entering the high-frequency wiring in steps of 0.1\,dB from the source power of the frequency extension module.

\section{Instrument Performance}

\begin{figure}
	\centering
		\includegraphics[width=1.00\columnwidth]{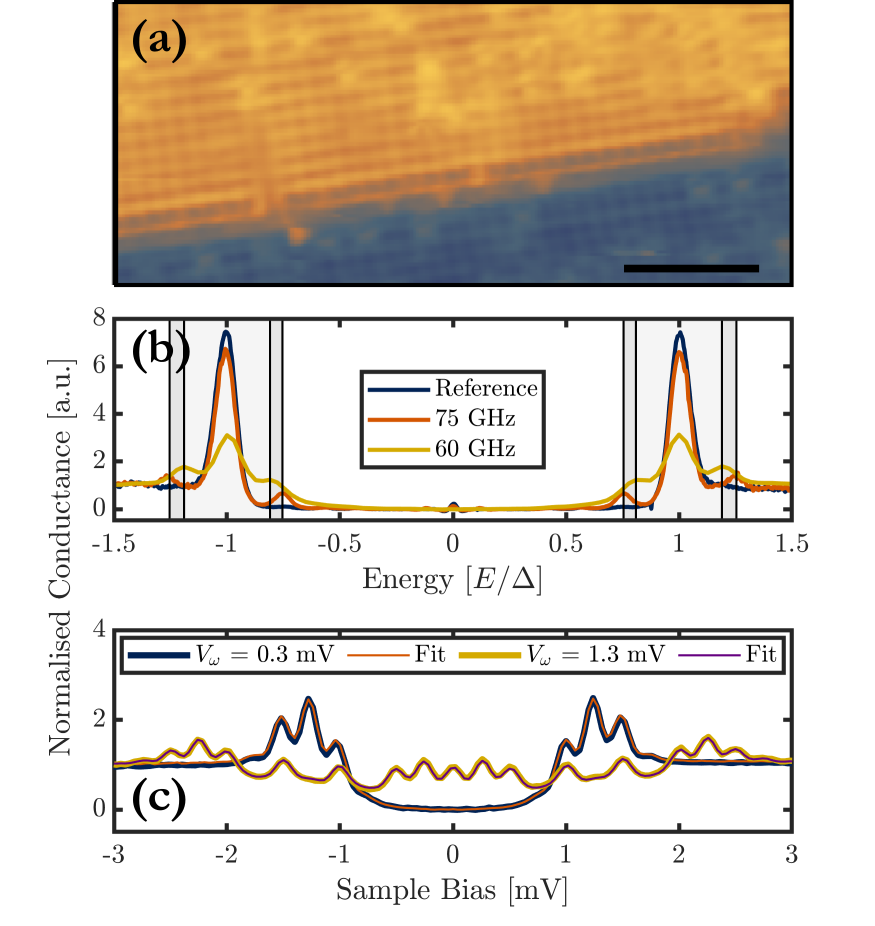}
	\caption{Performance test using a SIS tunnel junction. \textbf{(a)} Atomic resolution image of the oxygen reconstruction on a V(100) surface. \textbf{(b)} A typical reference spectrum of the SIS contact in absence of an RF signal (blue) is shown alongside conductance spectra under irradiation by a RF signal of 75\,GHz (orange) and 60\,GHz (yellow). Replica of the superconducting coherence peaks are clearly resolved. Their separation from the original position depends on the frequency of the RF signal. \textbf{(c)} Experimental data for a junction under irradiation by 60\,GHz radiation of varying power (blue and yellow) with fits from the model in Eq.\ \eqref{eq:QPtunnel} superimposed (orange and purple).}
	\label{fig:figure4}
\end{figure}

Key operational parameters of the instrument, such as base temperature, $z$-stability and $^3$He hold time, were unaffected by our modifications. We demonstrate the excellent STM performance in a series of test measurements on V(100). \figref{fig:figure4}(a) shows the oxygen reconstruction of the V(100) surface \cite{davies_surface_1980,foord_surfaces_1983} in atomic resolution at the base temperature of 310\,mK after the system upgrade. Tunnelling spectroscopy between a superconducting vanadium tip and the V(100) sample shows well-developed coherence peaks and a clear energy gap typical of superconductor-insulator-superconductor junctions (see \figref{fig:figure4}(b)) \cite{tinkham_superconductivity_1994,ast_sensing_2016,jack_critical_2016}.  

\begin{figure}
    \centering
    \includegraphics[width=1.00\columnwidth]{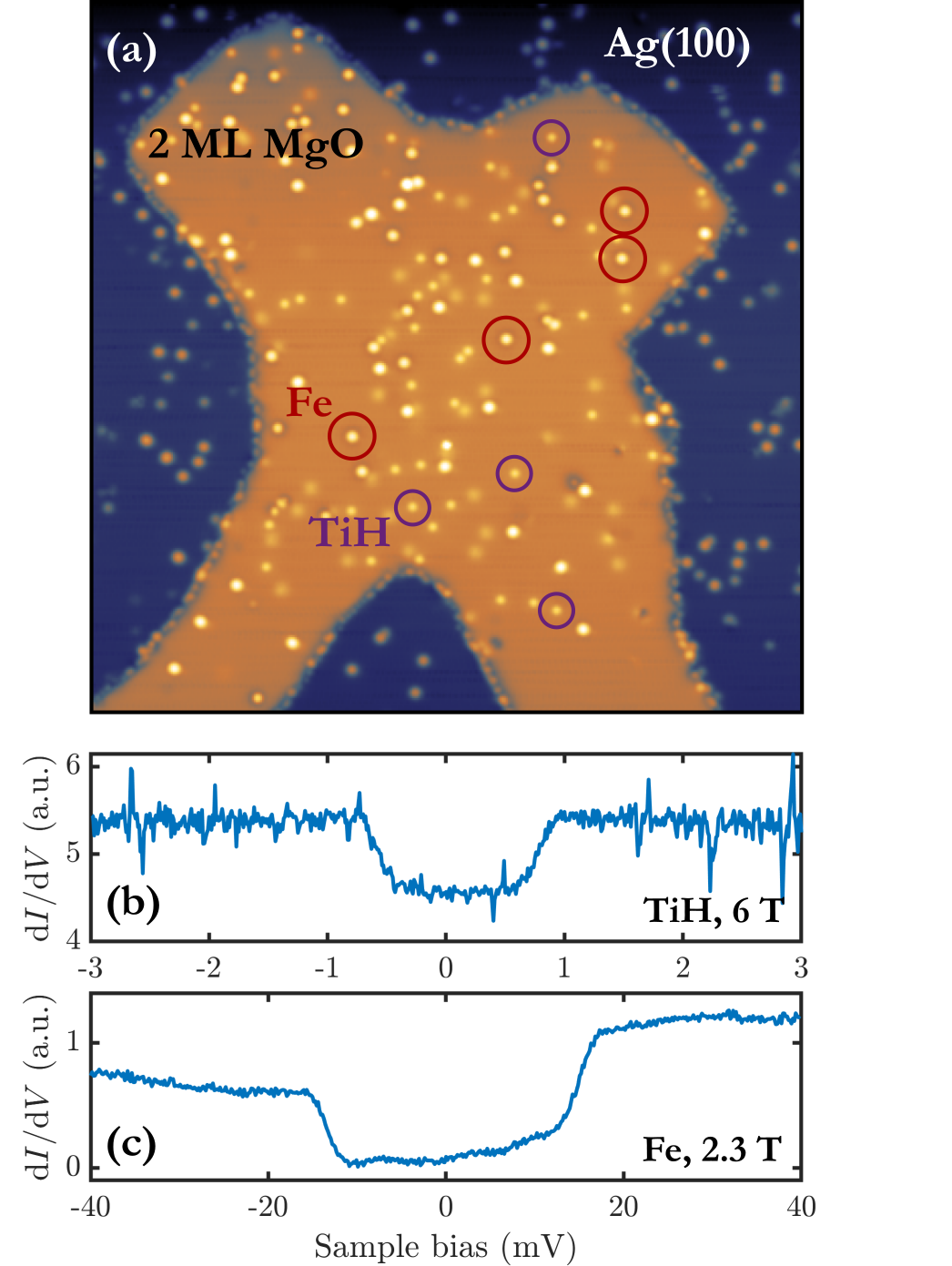}
    \caption{Individual atoms on a thin insulating layer for ESR-STM experiments. (a) A double layer MgO island on an Ag(100) surface serves as a decoupling layer for individual Fe atoms and TiH molecules. (b) IETS spectrum on a TiH molecule in 6 T out of plane external field. (c) IETS spectrum on an individual Fe atom in 2.3 T out of plane external field. The width of the excitation gap around zero bias is a fingerprint for the unambiguous identification of Fe and TiH.}
    \label{fig:figure5}
\end{figure}

\subsection{Microwave signal delivery}

We evaluate the performance of the microwave assembly by observing microwave-assisted tunnelling processes occurring in a superconductor-insulator-superconductor (SIS) junction between a V(100) sample and a vanadium tip driven by a high-frequency signal from the antenna. SIS junctions are well studied and a robust framework for data analysis is readily available \cite{falci_quasiparticle_1991,roychowdhury_microwave_2015,kot_microwave-assisted_2020,peters_resonant_2020}. The current in presence of a radio frequency signal is \cite{tien_multiphoton_1963,falci_quasiparticle_1991}:
\begin{equation}
	I_\text{QP}(V_{0},\,V_{\omega}) = \sum_{n=-\infty}^{\infty} J_n^2 \left( \frac{e V_{\omega}}{\hbar \omega} \right) I_{QP} \left(V_0 - \frac{n \hbar \omega}{e},0 \right),
	\label{eq:QPtunnel}
\end{equation}
where $I_\text{QP}(V,0)$ is the quasiparticle current in absence of any RF radiation, $V_{0}$ the DC bias voltage applied to the tunnel junction, $V_{\omega}$ the AC voltage dropping across the junction as a result of the RF signal, $J_n$ are the Bessel functions of the first kind of order $n$, and $e$ is the elementary charge. Further, $\hbar$ is the reduced Planck constant and $\omega = 2\pi\nu$, where $\nu$ is the microwave frequency. The sharp coherence peaks in the SIS tunnel data allow us to observe directly the effects of changing frequency. To analyse our data, we use reference spectra acquired using the same parameters as the corresponding microwave-assisted tunnelling spectrum, but with the RF signal switched off. This reference measurement is used as input for $I_{QP}$ in Eq.\ \eqref{eq:QPtunnel}. Fitting our model to the experimental data allows us to extract the frequency and amplitude of the RF signal arriving at the junction. All data shown is acquired at a base temperature of 310\,mK.

\figref{fig:figure4}(b) shows a typical conductance spectrum of the SIS junction in blue. We observe two prominent coherence peaks separated by twice the sum of the superconducting gaps of the tip and sample. At zero bias voltage, a small supercurrent flows through the junction as a result of the Josephson effect, proving that both tip and sample are in the superconducting state. The orange and yellow lines show conductance spectra of the SIS junction under the influence of RF radiation of 75\,GHz and 60\,GHz, respectively. Microwave driving of the junction gives rise to replica of the coherence peaks, offset by integer multiples of $h\nu$ from the main features and weighted by the square of the corresponding Bessel function.

We fit our data with the model given in Eq.\ \eqref{eq:QPtunnel} to extract the AC voltage dropping across the RF driven junction. \figref{fig:figure4}(c) shows sample fits for a junction under irradiation by 60\,GHz radio waves. As the RF power, and hence the AC voltage drop across the SIS junction, increase, higher order processes become visible. As the total current through the junction remains the same, microwave-assisted tunnelling leads to an overall decrease of the heights of individual peaks as the spectral weight is redistributed across a wide voltage range. The fitted curves are in excellent agreement with the data. Only a single frequency is needed in the model to reproduce the experimental results \cite{kot_microwave-assisted_2020}.

\begin{figure}
	\centering
		\includegraphics[width=1.00\columnwidth]{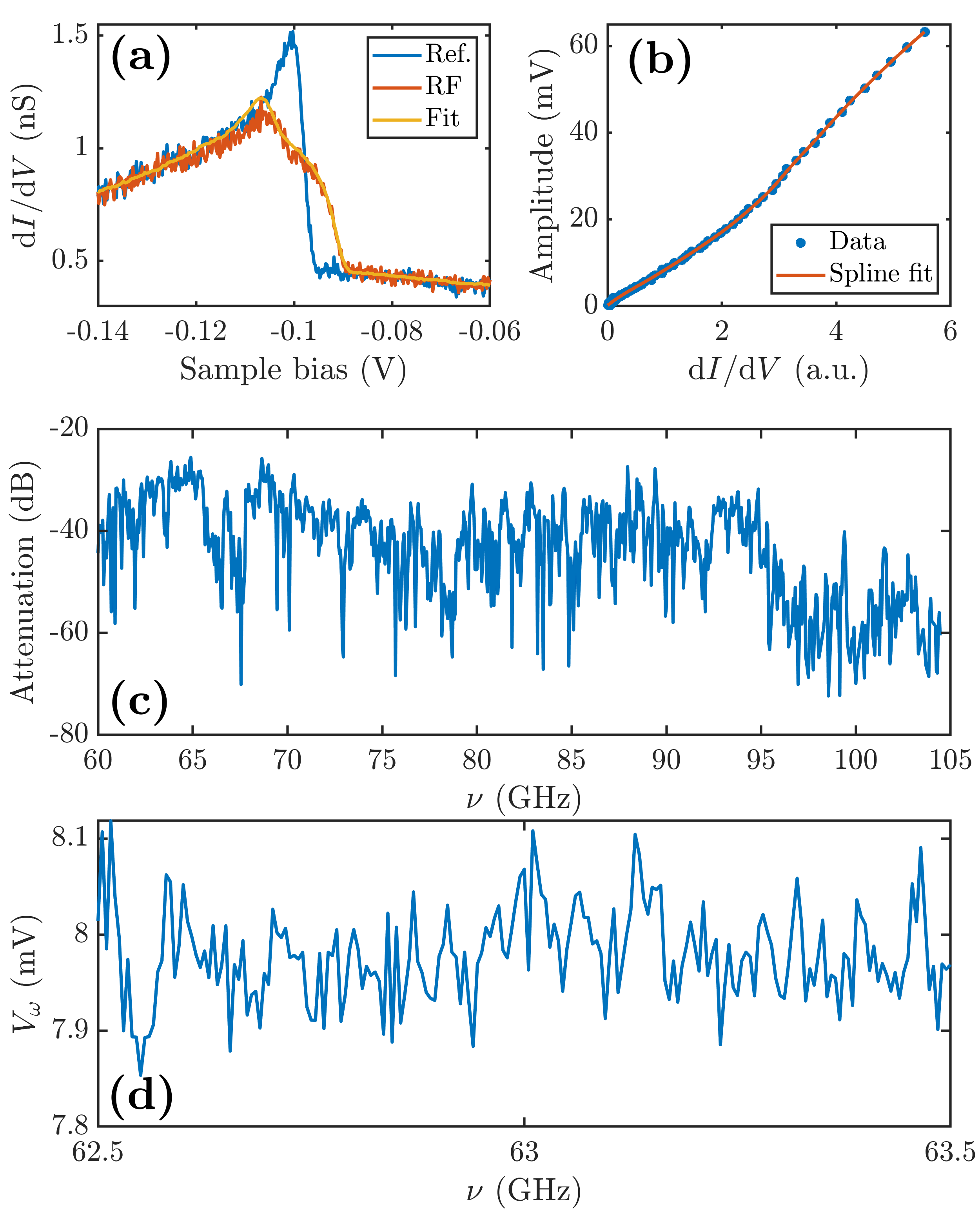}
		\caption{Generation of constant amplitude radio frequency sweeps at the tunnel junction. \textbf{(a)} Step-like feature measured on a TiH molecule (blue), spectrum under radio frequency irradiation (orange), and fit using Eq.\  \eqref{eq:QPtunnel} (yellow) with $\nu$=64.9\,GHz and $V_{\omega}$=7.5\,mV. The radio frequency amplitude determined from the fit is used to compute the transfer function. \textbf{(b)} Calibration curve mapping the measured lock-in signal to RF amplitude. The orange line is a smoothed spline interpolation, which we use to interpolate the calibration curve. \textbf{(c)} Transfer function measured from 60\,GHz to 105\,GHz. The operational range of the instrument extends far past the intended 60 to 90\,GHz. \textbf{(d)} Constant amplitude radio frequency sweep at 8\,mV setpoint value.}
	\label{fig:figure6}
\end{figure}

\begin{figure}
    \centering
    \includegraphics[width=1.00\columnwidth]{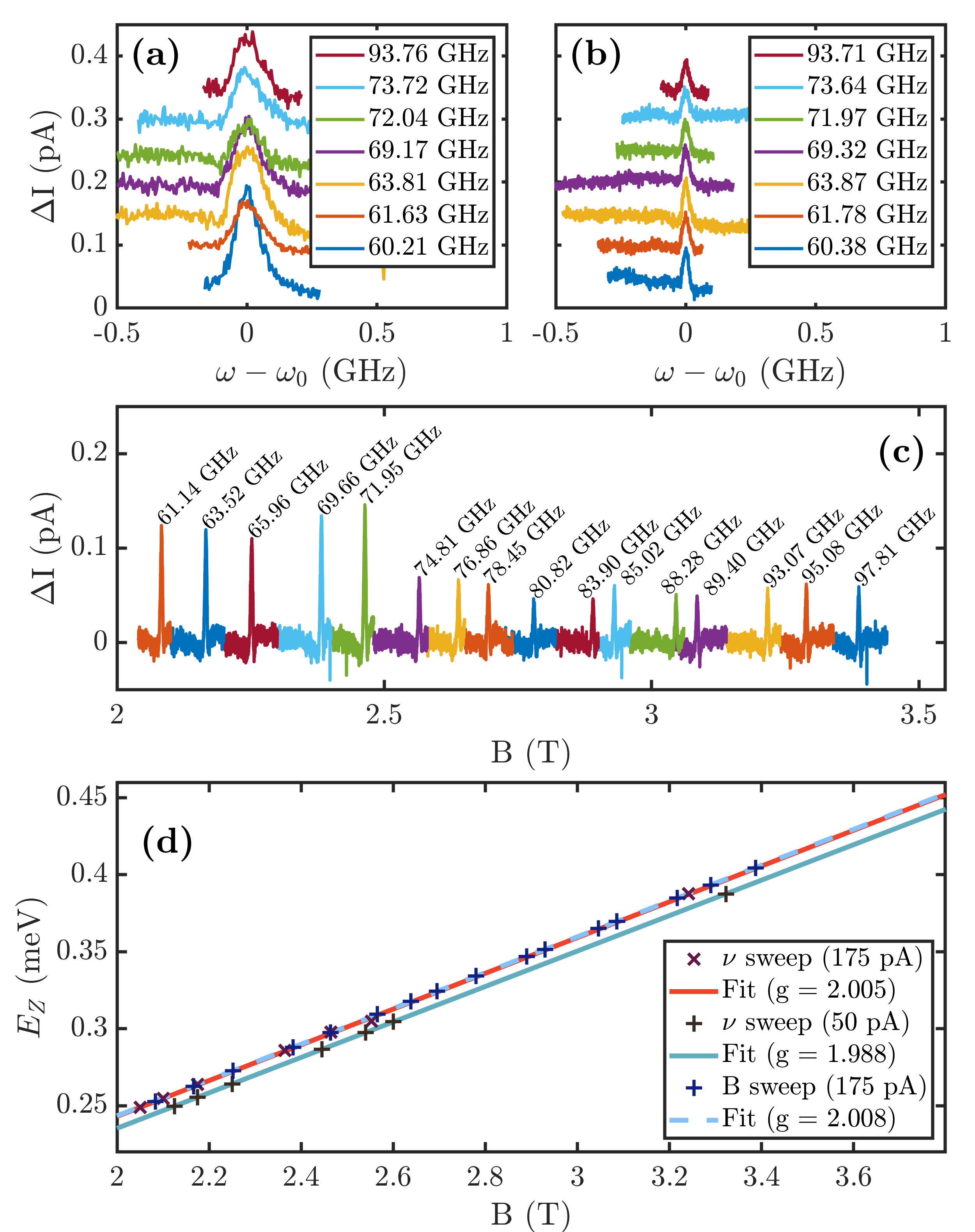}
    \caption{High-frequency ESR STM. ESR-STM signals from frequency sweeps at  175\,pA (a) and 50\,pA (b) setpoint current. The legend gives the centre frequency of the signal. The probe current is a source of decoherence that contributes to the broadening of the signal. (c) ESR-STM signals from field sweeps at  175\,pA setpoint current. The frequency of the excitations signal for each sweep is given in the annotations. (d) All ESR-STM measurements yield a $g$-factor close to 2.}
    \label{fig:figure7}
\end{figure}

\subsection{ESR-STM}

The prototypical spin systems for ESR-STM are individual metal atoms on an MgO decoupling layer on Ag(100) \cite{baumann_electron_2015}. We grow double-layer MgO islands by sublimation of Mg in a 1$\times$10$^{-6}$\,mbar oxygen atmosphere onto the sample held at 700\,K. Sublimation of Ti and Fe from an electron beam evaporator onto the cold sample (T$\lesssim$20\,K) yields individual Fe atoms and TiH molecules.  We use individual metal atoms on MgO/Ag(100) as a model system to provide proof of principle ESR-STM measurements. \figref{fig:figure5}(a) shows an STM topograph of a typical sample, once again demonstrating that the machine can resolve single atoms with ease. The tip is a sharpened PtIr wire, further prepared by field emission in the microscope and repeated indentation into the Ag(100). We identify individual Fe atoms and TiH molecules through their fingerprints in inelastic excitation tunnel spectroscopy (IETS), see \figref{fig:figure5}(b) and (c). 

The essential ingredient of any ESR measurement is the tuning of the spin system across its resonance with the microwave signal. This may be achieved through either changing the excitation frequency at fixed field ($\nu$ sweep), or the external field at fixed excitation frequency ($B$ sweep). In either case, it is necessary to calibrate the microwave amplitude in order to perform consistent and comparable measurements. Frequency sweeps further require the compensation of the antenna-junction transfer function to ensure a constant amplitude signal at throughout the measurement to suppress spurious signals.

We mainly follow Ref.\ \cite{paul_generation_2016} to calibrate the microwave signal generate constant amplitude radio frequency sweeps at the junction. A similar procedure is given in Ref.\ \cite{seifert_single-atom_2020}. Individual TiH molecules on MgO show a prominent step-like feature at a bias voltage of about $-80$\,mV (see blue spectrum in \figref{fig:figure6}(a)). We first determine the transfer function of the antenna assembly at one fixed frequency using Eq.\ \eqref{eq:QPtunnel}. We acquire a reference spectrum in absence of any RF signal and a tunnel spectrum under RF irradiation, as for the SIS case above. Then, we extract the signal amplitude $V_{\omega}$ through a fit (see yellow spectrum in \figref{fig:figure6}(a)) of the data for the irradiated junction (see red spectrum in  \figref{fig:figure6}(a)). The transfer function $T$ is then easily found from the known source amplitude $V_{\text{S}}$ according to 
\begin{equation}
	T = 20 \log (V_{\omega}/V_{\text{S}}).
\end{equation}
We then acquire a calibration curve by measuring the conductance near the centre of the slope of the TiH step as a function of source amplitude, see \figref{fig:figure6}(b). We use a smoothing cubic spline interpolation to fit the calibration function more accurately for arbitrary shapes. The smoothing parameter must be calculated such that the resulting smoothed spline can be inverted. This produces better fits than directly fitting to the inverted data. With the known transfer function value, these values can be converted into signal amplitudes at the junction. Since the TiH step is much  broader than $h\nu$ in the frequency range of interest, the frequency itself has no measurable effect on the RF spectrum and the calibration curve becomes a universal injective mapping of the conductance on $V_{\omega}$. 

By measuring the conductance of our tunnel junction with fixed signal attenuation while varying the source frequency and mapping the acquired values onto amplitudes using the inverted calibration curve, we measure the transfer function of the antenna assembly over a large frequency range, see \figref{fig:figure6}(c). 
The instrument performs well over the complete intended operational range of 60 to 90\,GHz and beyond. Losses do not become insurmountable until about 105\,GHz. Using the experimentally determined transfer function, we can compensate the losses in the antenna assembly through the adjustable attenuator to generate constant amplitude radio frequency sweeps at the tunnel junction, see \figref{fig:figure6}(d). We achieve signal amplitudes of 10\,mV throughout most of the frequency range, and up to about 90\,mV at select frequencies.

\subsection{ESR Signal}

We demonstrate the ESR capabilities of our setup using TiH molecules on double layer MgO/Ag(100) as a model system (see \figref{fig:figure5}). TiH adsorbed on the bridge site (between two oxygen atoms) of MgO is a \spinhalf\ system with a $g$-factor close to 2 in an out-of-plane field \cite{bae_enhanced_2018, yang_engineering_2017, steinbrecher_quantifying_2021, seifert_longitudinal_2020, seifert_single-atom_2020}. We attach one or more Fe atoms to the STM tip to generate a spin-polarised probe by picking them up from the surface \cite{baumann_electron_2015}. An out-of-plane magnetic field lifts the degeneracy between the spin states of the TiH molecule. In our frequency range, resonance is achieved in fields between about 2.15\,T and 3.75\,T. The Zeeman energy in these fields is about an order of magnitude larger than the thermal energy at the base temperature of 310\,mK and the spin systems are thermally initialised to their ground state. All of the data presented here are measured at a bias voltage of 100\,mV and the setpoint currents are referenced to this voltage.

The RF signal drives the resonant transition between the ground and excited states, resulting in a reduction of the spin-polarised signal on the atom as the time-averaged spin population is no longer thermal. By chopping the RF driving signal and locking in to the chopping frequency with a lock-in amplifier, we single out and record those parts of the signal that are affected by the driving, i.e. the ESR-STM signal \cite{baumann_electron_2015,paul_generation_2016}.

Our results are summarised in \figref{fig:figure7}. All data shown was acquired with the same spin-polarized microtip on the same TiH molecule adsorbed on a bridge site. Figure \ref{fig:figure7}(a) and (b) show frequency sweep ESR-STM measurements at 175\,pA and 50\,pA setpoint current, respectively. The RF amplitudes are 10\,mV and 15\,mV for the data at  175\,pA and 50\,pA, respectively. We find reasonably spaced intervals throughout the frequency band between 60\,GHz and 95\,GHz, where a $\nu$ sweep is possible (as indicated in the insets). As the probe current is itself a source of decoherence, the line width is substantially reduced at lower current. This reproduces the general trend that has been observed previously \cite{willke_probing_2018}. Also, the line widths of the measured ESR peaks are comparable to the published literature at lower frequencies ($\le40\,$GHz). 

\figref{fig:figure7}(c) shows ESR data from $B$ sweep measurements at sixteen different excitation frequencies spanning the available frequency range between 60\,GHz and 100\,GHz of the instrument. The microwave amplitude was set to 10\,mV. The baseline has been offset to zero for all sweeps. We find a fairly even distribution of suitable frequencies across a magnetic field range of almost 1.5\,T. Such a spread allows for a more accurate determination of the $g$-factor, which contains valuable information on the molecule and its environment \cite{steinbrecher_quantifying_2021}.

We calculate the corresponding Zeeman energies $E_\text{Z} = h\nu_{\text{res}}$ for all data sets, where $\nu_\text{res}$ is the resonance frequency. The $g$-factor of our spin system can then be extracted through a simple linear fit according to
\begin{equation}
    E_\text{Z} = 2\mu_\text{B} s g B, \label{eq:zeeman}
\end{equation}
where $\mu_\text{B}$ is the Bohr magneton and $s=1/2$ is the spin of the TiH molecule. The results are shown in Fig.\ \ref{fig:figure7}(d). We consistently obtain $g$-factors close to 2 for all our measurements, in agreement with previous results for TiH on the bridge site and in an out-of-plane field \cite{seifert_single-atom_2020}. The data points lie very well on the line given by Eq.\ \eqref{eq:zeeman} for both the $\nu$ sweep as well as the $B$ sweep, which were extracted from independent measurements at 175\,pA setpoint current shown in Fig.\ \ref{fig:figure7}(a) and (c). The fitted $g$-factors are $g=2.005$ ($\nu$ sweep) and $g=2.008$ ($B$ sweep), which lie within 0.15\%. For the $\nu$ sweep at 50\,pA setpoint current, we find a $g$-factor of $g=1.988$, which is slightly smaller than for the higher setpoint current. The offset between the $\nu$ sweeps at 50\,pA and 175\,pA setpoint current can be attributed to the different tip fields $B_\text{tip}$ felt by the TiH molecule. The tip fields are 95.7\,mT (B sweep) and 95.1\,mT ($\nu$ sweep) at 175\,pA current setpoint and 46.1\,mT for the 50\,pA current setpoint. The small changes in the $g$-factor can be attributed to tip-induced changes in the TiH bond length \cite{steinbrecher_quantifying_2021,seifert_longitudinal_2020}.

\section{Conclusion}

We have augmented a commercially available low-temperature STM system into a high-performance ESR-STM by the addition of a dedicated high-frequency line with antenna assembly to deliver radio frequency signals between 60 and 105\,GHz to the tunnel junction. Using commercially available components rated to high GHz frequencies, we achieve a high signal amplitude across a wide frequency range. The compensation of the transfer function allows us to keep the signal amplitude constant throughout our frequency range. In a series of proof-of-principle measurements, we measure an ESR signal on individual TiH molecules on a MgO decoupling layer in both frequency sweep and field sweep modes. In an operational field of several Tesla and at a base temperature of 310\,mK, excited state populations of typical \spinhalf\ systems at a resonance frequency above 60\,GHz are lower than 0.01\%. With these parameters, it becomes possible to study the intrinsic dynamics of individual spins with atomic resolution. Our approach can serve as a template to convert typical STMs into ESR-STMs capable of resolving the coherent dynamics of individual spins and magnetic nanostructures. 
\section{Acknowledgements}

We gratefully thank Andreas Heinrich, Klaus Kern, Aparajita Singha, Markus Ternes, and Philip Willke for fruitful discussions. This  work  was  funded  in  part  by  the  ERC  Consolidator Grant AbsoluteSpin (Grant No. 681164).


\begin{thebibliography}{33}%
\makeatletter
\providecommand \@ifxundefined [1]{%
 \@ifx{#1\undefined}
}%
\providecommand \@ifnum [1]{%
 \ifnum #1\expandafter \@firstoftwo
 \else \expandafter \@secondoftwo
 \fi
}%
\providecommand \@ifx [1]{%
 \ifx #1\expandafter \@firstoftwo
 \else \expandafter \@secondoftwo
 \fi
}%
\providecommand \natexlab [1]{#1}%
\providecommand \emph  [1]{``#1''}%
\providecommand \bibnamefont  [1]{#1}%
\providecommand \bibfnamefont [1]{#1}%
\providecommand \citenamefont [1]{#1}%
\providecommand \href@noop [0]{\@secondoftwo}%
\providecommand \href [0]{\begingroup \@sanitize@url \@href}%
\providecommand \@href[1]{\@@startlink{#1}\@@href}%
\providecommand \@@href[1]{\endgroup#1\@@endlink}%
\providecommand \@sanitize@url [0]{\catcode `\\12\catcode `\$12\catcode
  `\&12\catcode `\#12\catcode `\^12\catcode `\_12\catcode `\%12\relax}%
\providecommand \@@startlink[1]{}%
\providecommand \@@endlink[0]{}%
\providecommand \url  [0]{\begingroup\@sanitize@url \@url }%
\providecommand \@url [1]{\endgroup\@href {#1}{\urlprefix }}%
\providecommand \urlprefix  [0]{URL }%
\providecommand \Eprint [0]{\href }%
\providecommand \doibase [0]{http://dx.doi.org/}%
\providecommand \selectlanguage [0]{\@gobble}%
\providecommand \bibinfo  [0]{\@secondoftwo}%
\providecommand \bibfield  [0]{\@secondoftwo}%
\providecommand \translation [1]{[#1]}%
\providecommand \BibitemOpen [0]{}%
\providecommand \bibitemStop [0]{}%
\providecommand \bibitemNoStop [0]{.\EOS\space}%
\providecommand \EOS [0]{\spacefactor3000\relax}%
\providecommand \BibitemShut  [1]{\csname bibitem#1\endcsname}%
\let\auto@bib@innerbib\@empty
\bibitem [{\citenamefont {Heinrich}\ \emph {et~al.}(2004)\citenamefont
  {Heinrich}, \citenamefont {Gupta}, \citenamefont {Lutz},\ and\ \citenamefont
  {Eigler}}]{heinrich_single-atom_2004}%
  \BibitemOpen
  \bibfield  {author} {\bibinfo {author} {\bibfnamefont {A.~J.}\ \bibnamefont
  {Heinrich}}, \bibinfo {author} {\bibfnamefont {J.~A.}\ \bibnamefont {Gupta}},
  \bibinfo {author} {\bibfnamefont {C.~P.}\ \bibnamefont {Lutz}}, \ and\
  \bibinfo {author} {\bibfnamefont {D.~M.}\ \bibnamefont {Eigler}},\ }\bibfield
   {title} {{\selectlanguage {english}\emph {\bibinfo {title} {Single-atom
  spin-flip spectroscopy},}\ }}\href {\doibase 10.1126/science.1101077}
  {\bibfield  {journal} {\bibinfo  {journal} {Science}\ }\textbf {\bibinfo
  {volume} {306}},\ \bibinfo {pages} {466} (\bibinfo {year}
  {2004})}\BibitemShut {NoStop}%
\bibitem [{\citenamefont {Hirjibehedin}\ \emph {et~al.}(2007)\citenamefont
  {Hirjibehedin}, \citenamefont {Lin}, \citenamefont {Otte}, \citenamefont
  {Ternes}, \citenamefont {Lutz}, \citenamefont {Jones},\ and\ \citenamefont
  {Heinrich}}]{hirjibehedin_large_2007}%
  \BibitemOpen
  \bibfield  {author} {\bibinfo {author} {\bibfnamefont {C.~F.}\ \bibnamefont
  {Hirjibehedin}}, \bibinfo {author} {\bibfnamefont {C.-Y.}\ \bibnamefont
  {Lin}}, \bibinfo {author} {\bibfnamefont {A.~F.}\ \bibnamefont {Otte}},
  \bibinfo {author} {\bibfnamefont {M.}~\bibnamefont {Ternes}}, \bibinfo
  {author} {\bibfnamefont {C.~P.}\ \bibnamefont {Lutz}}, \bibinfo {author}
  {\bibfnamefont {B.~A.}\ \bibnamefont {Jones}}, \ and\ \bibinfo {author}
  {\bibfnamefont {A.~J.}\ \bibnamefont {Heinrich}},\ }\bibfield  {title}
  {{\selectlanguage {english}\emph {\bibinfo {title} {Large magnetic anisotropy
  of a single atomic spin embedded in a surface molecular network},}\ }}\href
  {\doibase 10.1126/science.1146110} {\bibfield  {journal} {\bibinfo  {journal}
  {Science}\ }\textbf {\bibinfo {volume} {317}},\ \bibinfo {pages} {1199}
  (\bibinfo {year} {2007})}\BibitemShut {NoStop}%
\bibitem [{\citenamefont {Baumann}\ \emph {et~al.}(2015)\citenamefont
  {Baumann}, \citenamefont {Paul}, \citenamefont {Choi}, \citenamefont {Lutz},
  \citenamefont {Ardavan},\ and\ \citenamefont
  {Heinrich}}]{baumann_electron_2015}%
  \BibitemOpen
  \bibfield  {author} {\bibinfo {author} {\bibfnamefont {S.}~\bibnamefont
  {Baumann}}, \bibinfo {author} {\bibfnamefont {W.}~\bibnamefont {Paul}},
  \bibinfo {author} {\bibfnamefont {T.}~\bibnamefont {Choi}}, \bibinfo {author}
  {\bibfnamefont {C.~P.}\ \bibnamefont {Lutz}}, \bibinfo {author}
  {\bibfnamefont {A.}~\bibnamefont {Ardavan}}, \ and\ \bibinfo {author}
  {\bibfnamefont {A.~J.}\ \bibnamefont {Heinrich}},\ }\bibfield  {title}
  {{\selectlanguage {english}\emph {\bibinfo {title} {Electron paramagnetic
  resonance of individual atoms on a surface},}\ }}\href {\doibase
  10.1126/science.aac8703} {\bibfield  {journal} {\bibinfo  {journal}
  {Science}\ }\textbf {\bibinfo {volume} {350}},\ \bibinfo {pages} {417}
  (\bibinfo {year} {2015})}\BibitemShut {NoStop}%
\bibitem [{\citenamefont {Willke}\ \emph {et~al.}(2018)\citenamefont {Willke},
  \citenamefont {Paul}, \citenamefont {Natterer}, \citenamefont {Yang},
  \citenamefont {Bae}, \citenamefont {Choi}, \citenamefont
  {Fernández-Rossier}, \citenamefont {Heinrich},\ and\ \citenamefont
  {Lutz}}]{willke_probing_2018}%
  \BibitemOpen
  \bibfield  {author} {\bibinfo {author} {\bibfnamefont {P.}~\bibnamefont
  {Willke}}, \bibinfo {author} {\bibfnamefont {W.}~\bibnamefont {Paul}},
  \bibinfo {author} {\bibfnamefont {F.~D.}\ \bibnamefont {Natterer}}, \bibinfo
  {author} {\bibfnamefont {K.}~\bibnamefont {Yang}}, \bibinfo {author}
  {\bibfnamefont {Y.}~\bibnamefont {Bae}}, \bibinfo {author} {\bibfnamefont
  {T.}~\bibnamefont {Choi}}, \bibinfo {author} {\bibfnamefont {J.}~\bibnamefont
  {Fernández-Rossier}}, \bibinfo {author} {\bibfnamefont {A.~J.}\ \bibnamefont
  {Heinrich}}, \ and\ \bibinfo {author} {\bibfnamefont {C.~P.}\ \bibnamefont
  {Lutz}},\ }\bibfield  {title} {{\selectlanguage {english}\emph {\bibinfo
  {title} {Probing quantum coherence in single-atom electron spin resonance},}\
  }}\href {\doibase 10.1126/sciadv.aaq1543} {\bibfield  {journal} {\bibinfo
  {journal} {Science Advances}\ }\textbf {\bibinfo {volume} {4}},\ \bibinfo
  {pages} {eaaq1543} (\bibinfo {year} {2018})}\BibitemShut {NoStop}%
\bibitem [{\citenamefont {Yang}\ \emph {et~al.}(2019)\citenamefont {Yang},
  \citenamefont {Paul}, \citenamefont {Phark}, \citenamefont {Willke},
  \citenamefont {Bae}, \citenamefont {Choi}, \citenamefont {Esat},
  \citenamefont {Ardavan}, \citenamefont {Heinrich},\ and\ \citenamefont
  {Lutz}}]{yang_coherent_2019}%
  \BibitemOpen
  \bibfield  {author} {\bibinfo {author} {\bibfnamefont {K.}~\bibnamefont
  {Yang}}, \bibinfo {author} {\bibfnamefont {W.}~\bibnamefont {Paul}}, \bibinfo
  {author} {\bibfnamefont {S.-H.}\ \bibnamefont {Phark}}, \bibinfo {author}
  {\bibfnamefont {P.}~\bibnamefont {Willke}}, \bibinfo {author} {\bibfnamefont
  {Y.}~\bibnamefont {Bae}}, \bibinfo {author} {\bibfnamefont {T.}~\bibnamefont
  {Choi}}, \bibinfo {author} {\bibfnamefont {T.}~\bibnamefont {Esat}}, \bibinfo
  {author} {\bibfnamefont {A.}~\bibnamefont {Ardavan}}, \bibinfo {author}
  {\bibfnamefont {A.~J.}\ \bibnamefont {Heinrich}}, \ and\ \bibinfo {author}
  {\bibfnamefont {C.~P.}\ \bibnamefont {Lutz}},\ }\bibfield  {title}
  {{\selectlanguage {english}\emph {\bibinfo {title} {Coherent spin
  manipulation of individual atoms on a surface},}\ }}\href {\doibase
  10.1126/science.aay6779} {\bibfield  {journal} {\bibinfo  {journal}
  {Science}\ }\textbf {\bibinfo {volume} {366}},\ \bibinfo {pages} {509}
  (\bibinfo {year} {2019})}\BibitemShut {NoStop}%
\bibitem [{\citenamefont {Seifert}\ \emph
  {et~al.}(2020{\natexlab{a}})\citenamefont {Seifert}, \citenamefont {Kovarik},
  \citenamefont {Nistor}, \citenamefont {Persichetti}, \citenamefont
  {Stepanow},\ and\ \citenamefont {Gambardella}}]{seifert_single-atom_2020}%
  \BibitemOpen
  \bibfield  {author} {\bibinfo {author} {\bibfnamefont {T.~S.}\ \bibnamefont
  {Seifert}}, \bibinfo {author} {\bibfnamefont {S.}~\bibnamefont {Kovarik}},
  \bibinfo {author} {\bibfnamefont {C.}~\bibnamefont {Nistor}}, \bibinfo
  {author} {\bibfnamefont {L.}~\bibnamefont {Persichetti}}, \bibinfo {author}
  {\bibfnamefont {S.}~\bibnamefont {Stepanow}}, \ and\ \bibinfo {author}
  {\bibfnamefont {P.}~\bibnamefont {Gambardella}},\ }\bibfield  {title} {\emph
  {\bibinfo {title} {Single-atom electron paramagnetic resonance in a scanning
  tunneling microscope driven by a radio-frequency antenna at 4 {K}},}\ }\href
  {\doibase 10.1103/PhysRevResearch.2.013032} {\bibfield  {journal} {\bibinfo
  {journal} {Physical Review Research}\ }\textbf {\bibinfo {volume} {2}},\
  \bibinfo {pages} {013032} (\bibinfo {year} {2020}{\natexlab{a}})}\BibitemShut
  {NoStop}%
\bibitem [{\citenamefont {Veldman}\ \emph {et~al.}(2021)\citenamefont
  {Veldman}, \citenamefont {Farinacci}, \citenamefont {Rejali}, \citenamefont
  {Broekhoven}, \citenamefont {Gobeil}, \citenamefont {Coffey}, \citenamefont
  {Ternes},\ and\ \citenamefont {Otte}}]{veldman_free_2021}%
  \BibitemOpen
  \bibfield  {author} {\bibinfo {author} {\bibfnamefont {L.~M.}\ \bibnamefont
  {Veldman}}, \bibinfo {author} {\bibfnamefont {L.}~\bibnamefont {Farinacci}},
  \bibinfo {author} {\bibfnamefont {R.}~\bibnamefont {Rejali}}, \bibinfo
  {author} {\bibfnamefont {R.}~\bibnamefont {Broekhoven}}, \bibinfo {author}
  {\bibfnamefont {J.}~\bibnamefont {Gobeil}}, \bibinfo {author} {\bibfnamefont
  {D.}~\bibnamefont {Coffey}}, \bibinfo {author} {\bibfnamefont
  {M.}~\bibnamefont {Ternes}}, \ and\ \bibinfo {author} {\bibfnamefont {A.~F.}\
  \bibnamefont {Otte}},\ }\bibfield  {title} {{\selectlanguage {english}\emph
  {\bibinfo {title} {Free coherent evolution of a coupled atomic spin system
  initialized by electron scattering},}\ }}\href
  {https://www.science.org/doi/abs/10.1126/science.abg8223} {\bibfield
  {journal} {\bibinfo  {journal} {Science}\ }\textbf {\bibinfo {volume}
  {372}},\ \bibinfo {pages} {964} (\bibinfo {year} {2021})}\BibitemShut
  {NoStop}%
\bibitem [{\citenamefont {Balatsky}\ \emph {et~al.}(2012)\citenamefont
  {Balatsky}, \citenamefont {Nishijima},\ and\ \citenamefont
  {Manassen}}]{balatsky_electron_2012}%
  \BibitemOpen
  \bibfield  {author} {\bibinfo {author} {\bibfnamefont {A.~V.}\ \bibnamefont
  {Balatsky}}, \bibinfo {author} {\bibfnamefont {M.}~\bibnamefont {Nishijima}},
  \ and\ \bibinfo {author} {\bibfnamefont {Y.}~\bibnamefont {Manassen}},\
  }\bibfield  {title} {\emph {\bibinfo {title} {Electron spin
  resonance-scanning tunneling microscopy},}\ }\href {\doibase
  10.1080/00018732.2012.668775} {\bibfield  {journal} {\bibinfo  {journal}
  {Advances in Physics}\ }\textbf {\bibinfo {volume} {61}},\ \bibinfo {pages}
  {117} (\bibinfo {year} {2012})}\BibitemShut {NoStop}%
\bibitem [{\citenamefont {Müllegger}\ \emph {et~al.}(2014)\citenamefont
  {Müllegger}, \citenamefont {Tebi}, \citenamefont {Das}, \citenamefont
  {Schöffberger}, \citenamefont {Faschinger},\ and\ \citenamefont
  {Koch}}]{mullegger_radio_2014}%
  \BibitemOpen
  \bibfield  {author} {\bibinfo {author} {\bibfnamefont {S.}~\bibnamefont
  {Müllegger}}, \bibinfo {author} {\bibfnamefont {S.}~\bibnamefont {Tebi}},
  \bibinfo {author} {\bibfnamefont {A.~K.}\ \bibnamefont {Das}}, \bibinfo
  {author} {\bibfnamefont {W.}~\bibnamefont {Schöffberger}}, \bibinfo {author}
  {\bibfnamefont {F.}~\bibnamefont {Faschinger}}, \ and\ \bibinfo {author}
  {\bibfnamefont {R.}~\bibnamefont {Koch}},\ }\bibfield  {title} {\emph
  {\bibinfo {title} {Radio frequency scanning tunneling spectroscopy for
  single-molecule spin resonance},}\ }\href {\doibase
  10.1103/PhysRevLett.113.133001} {\bibfield  {journal} {\bibinfo  {journal}
  {Phys. Rev. Lett.}\ }\textbf {\bibinfo {volume} {113}},\ \bibinfo {pages}
  {133001} (\bibinfo {year} {2014})}\BibitemShut {NoStop}%
\bibitem [{\citenamefont {Binnig}\ \emph {et~al.}(1982)\citenamefont {Binnig},
  \citenamefont {Rohrer}, \citenamefont {Gerber},\ and\ \citenamefont
  {Weibel}}]{binnig_surface_1982}%
  \BibitemOpen
  \bibfield  {author} {\bibinfo {author} {\bibfnamefont {G.}~\bibnamefont
  {Binnig}}, \bibinfo {author} {\bibfnamefont {H.}~\bibnamefont {Rohrer}},
  \bibinfo {author} {\bibfnamefont {C.}~\bibnamefont {Gerber}}, \ and\ \bibinfo
  {author} {\bibfnamefont {E.}~\bibnamefont {Weibel}},\ }\bibfield  {title}
  {\emph {\bibinfo {title} {Surface {Studies} by {Scanning} {Tunneling}
  {Microscopy}},}\ }\href {\doibase 10.1103/PhysRevLett.49.57} {\bibfield
  {journal} {\bibinfo  {journal} {Physical Review Letters}\ }\textbf {\bibinfo
  {volume} {49}},\ \bibinfo {pages} {57} (\bibinfo {year} {1982})}\BibitemShut
  {NoStop}%
\bibitem [{\citenamefont {Loth}\ \emph {et~al.}(2010)\citenamefont {Loth},
  \citenamefont {Etzkorn}, \citenamefont {Lutz}, \citenamefont {Eigler},\ and\
  \citenamefont {Heinrich}}]{loth_measurement_2010}%
  \BibitemOpen
  \bibfield  {author} {\bibinfo {author} {\bibfnamefont {S.}~\bibnamefont
  {Loth}}, \bibinfo {author} {\bibfnamefont {M.}~\bibnamefont {Etzkorn}},
  \bibinfo {author} {\bibfnamefont {C.~P.}\ \bibnamefont {Lutz}}, \bibinfo
  {author} {\bibfnamefont {D.~M.}\ \bibnamefont {Eigler}}, \ and\ \bibinfo
  {author} {\bibfnamefont {A.~J.}\ \bibnamefont {Heinrich}},\ }\bibfield
  {title} {{\selectlanguage {english}\emph {\bibinfo {title} {Measurement of
  fast electron spin relaxation times with atomic resolution},}\ }}\href
  {\doibase 10.1126/science.1191688} {\bibfield  {journal} {\bibinfo  {journal}
  {Science}\ }\textbf {\bibinfo {volume} {329}},\ \bibinfo {pages} {1628}
  (\bibinfo {year} {2010})}\BibitemShut {NoStop}%
\bibitem [{\citenamefont {Cocker}\ \emph {et~al.}(2013)\citenamefont {Cocker},
  \citenamefont {Jelic}, \citenamefont {Gupta}, \citenamefont {Molesky},
  \citenamefont {Burgess}, \citenamefont {Reyes}, \citenamefont {Titova},
  \citenamefont {Tsui}, \citenamefont {Freeman},\ and\ \citenamefont
  {Hegmann}}]{cocker_ultrafast_2013}%
  \BibitemOpen
  \bibfield  {author} {\bibinfo {author} {\bibfnamefont {T.~L.}\ \bibnamefont
  {Cocker}}, \bibinfo {author} {\bibfnamefont {V.}~\bibnamefont {Jelic}},
  \bibinfo {author} {\bibfnamefont {M.}~\bibnamefont {Gupta}}, \bibinfo
  {author} {\bibfnamefont {S.~J.}\ \bibnamefont {Molesky}}, \bibinfo {author}
  {\bibfnamefont {J.~A.~J.}\ \bibnamefont {Burgess}}, \bibinfo {author}
  {\bibfnamefont {G.~D.~L.}\ \bibnamefont {Reyes}}, \bibinfo {author}
  {\bibfnamefont {L.~V.}\ \bibnamefont {Titova}}, \bibinfo {author}
  {\bibfnamefont {Y.~Y.}\ \bibnamefont {Tsui}}, \bibinfo {author}
  {\bibfnamefont {M.~R.}\ \bibnamefont {Freeman}}, \ and\ \bibinfo {author}
  {\bibfnamefont {F.~A.}\ \bibnamefont {Hegmann}},\ }\bibfield  {title}
  {{\selectlanguage {english}\emph {\bibinfo {title} {An ultrafast terahertz
  scanning tunnelling microscope},}\ }}\href {\doibase
  10.1038/nphoton.2013.151} {\bibfield  {journal} {\bibinfo  {journal} {Nat
  Photon}\ }\textbf {\bibinfo {volume} {7}},\ \bibinfo {pages} {620} (\bibinfo
  {year} {2013})}\BibitemShut {NoStop}%
\bibitem [{\citenamefont {Yoshida}\ \emph {et~al.}(2014)\citenamefont
  {Yoshida}, \citenamefont {Aizawa}, \citenamefont {Wang}, \citenamefont
  {Oshima}, \citenamefont {Mera}, \citenamefont {Matsuyama}, \citenamefont
  {Oigawa}, \citenamefont {Takeuchi},\ and\ \citenamefont
  {Shigekawa}}]{yoshida_probing_2014}%
  \BibitemOpen
  \bibfield  {author} {\bibinfo {author} {\bibfnamefont {S.}~\bibnamefont
  {Yoshida}}, \bibinfo {author} {\bibfnamefont {Y.}~\bibnamefont {Aizawa}},
  \bibinfo {author} {\bibfnamefont {Z.-h.}\ \bibnamefont {Wang}}, \bibinfo
  {author} {\bibfnamefont {R.}~\bibnamefont {Oshima}}, \bibinfo {author}
  {\bibfnamefont {Y.}~\bibnamefont {Mera}}, \bibinfo {author} {\bibfnamefont
  {E.}~\bibnamefont {Matsuyama}}, \bibinfo {author} {\bibfnamefont
  {H.}~\bibnamefont {Oigawa}}, \bibinfo {author} {\bibfnamefont
  {O.}~\bibnamefont {Takeuchi}}, \ and\ \bibinfo {author} {\bibfnamefont
  {H.}~\bibnamefont {Shigekawa}},\ }\bibfield  {title} {{\selectlanguage
  {english}\emph {\bibinfo {title} {Probing ultrafast spin dynamics with
  optical pump–probe scanning tunnelling microscopy},}\ }}\href {\doibase
  10.1038/nnano.2014.125} {\bibfield  {journal} {\bibinfo  {journal} {Nature
  Nanotechnology}\ }\textbf {\bibinfo {volume} {9}},\ \bibinfo {pages} {588}
  (\bibinfo {year} {2014})}\BibitemShut {NoStop}%
\bibitem [{\citenamefont {Gutzler}\ \emph {et~al.}(2021)\citenamefont
  {Gutzler}, \citenamefont {Garg}, \citenamefont {Ast}, \citenamefont
  {Kuhnke},\ and\ \citenamefont {Kern}}]{gutzler_light-matter_2021}%
  \BibitemOpen
  \bibfield  {author} {\bibinfo {author} {\bibfnamefont {R.}~\bibnamefont
  {Gutzler}}, \bibinfo {author} {\bibfnamefont {M.}~\bibnamefont {Garg}},
  \bibinfo {author} {\bibfnamefont {C.~R.}\ \bibnamefont {Ast}}, \bibinfo
  {author} {\bibfnamefont {K.}~\bibnamefont {Kuhnke}}, \ and\ \bibinfo {author}
  {\bibfnamefont {K.}~\bibnamefont {Kern}},\ }\bibfield  {title}
  {{\selectlanguage {english}\emph {\bibinfo {title} {Light–matter
  interaction at atomic scales},}\ }}\href {\doibase
  10.1038/s42254-021-00306-5} {\bibfield  {journal} {\bibinfo  {journal}
  {Nature Reviews Physics}\ ,\ \bibinfo {pages} {1}} (\bibinfo {year}
  {2021})}\BibitemShut {NoStop}%
\bibitem [{\citenamefont {Paul}\ \emph {et~al.}(2016)\citenamefont {Paul},
  \citenamefont {Baumann}, \citenamefont {Lutz},\ and\ \citenamefont
  {Heinrich}}]{paul_generation_2016}%
  \BibitemOpen
  \bibfield  {author} {\bibinfo {author} {\bibfnamefont {W.}~\bibnamefont
  {Paul}}, \bibinfo {author} {\bibfnamefont {S.}~\bibnamefont {Baumann}},
  \bibinfo {author} {\bibfnamefont {C.~P.}\ \bibnamefont {Lutz}}, \ and\
  \bibinfo {author} {\bibfnamefont {A.~J.}\ \bibnamefont {Heinrich}},\
  }\bibfield  {title} {\emph {\bibinfo {title} {Generation of
  constant-amplitude radio-frequency sweeps at a tunnel junction for spin
  resonance {STM}},}\ }\href {\doibase 10.1063/1.4955446} {\bibfield  {journal}
  {\bibinfo  {journal} {Review of Scientific Instruments}\ }\textbf {\bibinfo
  {volume} {87}},\ \bibinfo {pages} {074703} (\bibinfo {year}
  {2016})}\BibitemShut {NoStop}%
\bibitem [{\citenamefont {Natterer}\ \emph {et~al.}(2019)\citenamefont
  {Natterer}, \citenamefont {Patthey}, \citenamefont {Bilgeri}, \citenamefont
  {Forrester}, \citenamefont {Weiss},\ and\ \citenamefont
  {Brune}}]{natterer_upgrade_2019}%
  \BibitemOpen
  \bibfield  {author} {\bibinfo {author} {\bibfnamefont {F.~D.}\ \bibnamefont
  {Natterer}}, \bibinfo {author} {\bibfnamefont {F.}~\bibnamefont {Patthey}},
  \bibinfo {author} {\bibfnamefont {T.}~\bibnamefont {Bilgeri}}, \bibinfo
  {author} {\bibfnamefont {P.~R.}\ \bibnamefont {Forrester}}, \bibinfo {author}
  {\bibfnamefont {N.}~\bibnamefont {Weiss}}, \ and\ \bibinfo {author}
  {\bibfnamefont {H.}~\bibnamefont {Brune}},\ }\bibfield  {title} {\emph
  {\bibinfo {title} {Upgrade of a low-temperature scanning tunneling microscope
  for electron-spin resonance},}\ }\href {\doibase 10.1063/1.5065384}
  {\bibfield  {journal} {\bibinfo  {journal} {Review of Scientific
  Instruments}\ }\textbf {\bibinfo {volume} {90}},\ \bibinfo {pages} {013706}
  (\bibinfo {year} {2019})},\ \bibinfo {note} {publisher: American Institute of
  Physics}\BibitemShut {NoStop}%
\bibitem [{\citenamefont {Friedlein}\ \emph {et~al.}(2019)\citenamefont
  {Friedlein}, \citenamefont {Harm}, \citenamefont {Lindner}, \citenamefont
  {Bargsten}, \citenamefont {Bazarnik}, \citenamefont {Krause},\ and\
  \citenamefont {Wiesendanger}}]{friedlein_radio-frequency_2019}%
  \BibitemOpen
  \bibfield  {author} {\bibinfo {author} {\bibfnamefont {J.}~\bibnamefont
  {Friedlein}}, \bibinfo {author} {\bibfnamefont {J.}~\bibnamefont {Harm}},
  \bibinfo {author} {\bibfnamefont {P.}~\bibnamefont {Lindner}}, \bibinfo
  {author} {\bibfnamefont {L.}~\bibnamefont {Bargsten}}, \bibinfo {author}
  {\bibfnamefont {M.}~\bibnamefont {Bazarnik}}, \bibinfo {author}
  {\bibfnamefont {S.}~\bibnamefont {Krause}}, \ and\ \bibinfo {author}
  {\bibfnamefont {R.}~\bibnamefont {Wiesendanger}},\ }\bibfield  {title} {\emph
  {\bibinfo {title} {A radio-frequency spin-polarized scanning tunneling
  microscope},}\ }\href {\doibase 10.1063/1.5104317} {\bibfield  {journal}
  {\bibinfo  {journal} {Review of Scientific Instruments}\ }\textbf {\bibinfo
  {volume} {90}},\ \bibinfo {pages} {123705} (\bibinfo {year} {2019})},\
  \bibinfo {note} {publisher: American Institute of Physics}\BibitemShut
  {NoStop}%
\bibitem [{\citenamefont {van Weerdenburg}\ \emph {et~al.}(2021)\citenamefont
  {van Weerdenburg}, \citenamefont {Steinbrecher}, \citenamefont {van
  Mullekom}, \citenamefont {Gerritsen}, \citenamefont {von Allwörden},
  \citenamefont {Natterer},\ and\ \citenamefont
  {Khajetoorians}}]{weerdenburg_scanning_2021}%
  \BibitemOpen
  \bibfield  {author} {\bibinfo {author} {\bibfnamefont {W.~M.~J.}\
  \bibnamefont {van Weerdenburg}}, \bibinfo {author} {\bibfnamefont
  {M.}~\bibnamefont {Steinbrecher}}, \bibinfo {author} {\bibfnamefont
  {N.~P.~E.}\ \bibnamefont {van Mullekom}}, \bibinfo {author} {\bibfnamefont
  {J.~W.}\ \bibnamefont {Gerritsen}}, \bibinfo {author} {\bibfnamefont
  {H.}~\bibnamefont {von Allwörden}}, \bibinfo {author} {\bibfnamefont
  {F.~D.}\ \bibnamefont {Natterer}}, \ and\ \bibinfo {author} {\bibfnamefont
  {A.~A.}\ \bibnamefont {Khajetoorians}},\ }\bibfield  {title} {\emph {\bibinfo
  {title} {A scanning tunneling microscope capable of electron spin resonance
  and pump–probe spectroscopy at {mK} temperature and in vector magnetic
  field},}\ }\href {\doibase 10.1063/5.0040011} {\bibfield  {journal} {\bibinfo
   {journal} {Review of Scientific Instruments}\ }\textbf {\bibinfo {volume}
  {92}},\ \bibinfo {pages} {033906} (\bibinfo {year} {2021})}\BibitemShut
  {NoStop}%
\bibitem [{\citenamefont {Kot}\ \emph {et~al.}(2020)\citenamefont {Kot},
  \citenamefont {Drost}, \citenamefont {Uhl}, \citenamefont {Ankerhold},
  \citenamefont {Cuevas},\ and\ \citenamefont
  {Ast}}]{kot_microwave-assisted_2020}%
  \BibitemOpen
  \bibfield  {author} {\bibinfo {author} {\bibfnamefont {P.}~\bibnamefont
  {Kot}}, \bibinfo {author} {\bibfnamefont {R.}~\bibnamefont {Drost}}, \bibinfo
  {author} {\bibfnamefont {M.}~\bibnamefont {Uhl}}, \bibinfo {author}
  {\bibfnamefont {J.}~\bibnamefont {Ankerhold}}, \bibinfo {author}
  {\bibfnamefont {J.~C.}\ \bibnamefont {Cuevas}}, \ and\ \bibinfo {author}
  {\bibfnamefont {C.~R.}\ \bibnamefont {Ast}},\ }\bibfield  {title} {\emph
  {\bibinfo {title} {Microwave-assisted tunneling and interference effects in
  superconducting junctions under fast driving signals},}\ }\href {\doibase
  10.1103/PhysRevB.101.134507} {\bibfield  {journal} {\bibinfo  {journal}
  {Physical Review B}\ }\textbf {\bibinfo {volume} {101}},\ \bibinfo {pages}
  {134507} (\bibinfo {year} {2020})}\BibitemShut {NoStop}%
\bibitem [{\citenamefont {Tinkham}(1996)}]{tinkham_superconductivity_1994}%
  \BibitemOpen
  \bibfield  {author} {\bibinfo {author} {\bibfnamefont {M.}~\bibnamefont
  {Tinkham}},\ }\href {http://store.doverpublications.com/0486435032.html}
  {\emph {\bibinfo {title} {Introduction to {Superconductivity}}}}\ (\bibinfo
  {publisher} {Dover},\ \bibinfo {year} {1996})\BibitemShut {NoStop}%
\bibitem [{\citenamefont {Merkt}()}]{merkt_entwurf_2017}%
  \BibitemOpen
  \bibfield  {author} {\bibinfo {author} {\bibfnamefont {J.}~\bibnamefont
  {Merkt}},\ }\href@noop {} {\emph {\bibinfo {title} {{Entwurf und Untersuchung
  einer Antenne für 84GHz zur Strahlungseinkopplung in ein STM}},}\ }\bibinfo
  {note} {{}Bachelor's Thesis, Karlsruher Institut für Technologie,
  (2017)}\BibitemShut {NoStop}%
\bibitem [{\citenamefont {Davies}\ and\ \citenamefont
  {Lambert}(1980)}]{davies_surface_1980}%
  \BibitemOpen
  \bibfield  {author} {\bibinfo {author} {\bibfnamefont {P.~W.}\ \bibnamefont
  {Davies}}\ and\ \bibinfo {author} {\bibfnamefont {R.~M.}\ \bibnamefont
  {Lambert}},\ }\bibfield  {title} {{\selectlanguage {english}\emph {\bibinfo
  {title} {Surface chemistry of the metal-halogen interface: {Bromine}
  chemisorption and related studies on vanadium (100)},}\ }}\href {\doibase
  10.1016/0039-6028(80)90196-X} {\bibfield  {journal} {\bibinfo  {journal}
  {Surface Science}\ }\textbf {\bibinfo {volume} {95}},\ \bibinfo {pages} {571}
  (\bibinfo {year} {1980})}\BibitemShut {NoStop}%
\bibitem [{\citenamefont {Foord}\ \emph {et~al.}(1983)\citenamefont {Foord},
  \citenamefont {Reed},\ and\ \citenamefont {Lambert}}]{foord_surfaces_1983}%
  \BibitemOpen
  \bibfield  {author} {\bibinfo {author} {\bibfnamefont {J.~S.}\ \bibnamefont
  {Foord}}, \bibinfo {author} {\bibfnamefont {A.~P.~C.}\ \bibnamefont {Reed}},
  \ and\ \bibinfo {author} {\bibfnamefont {R.~M.}\ \bibnamefont {Lambert}},\
  }\bibfield  {title} {{\selectlanguage {english}\emph {\bibinfo {title} {The
  (100) surfaces of chromium and vanadium: {Reconsiderations} of their
  structure and reactivity},}\ }}\href {\doibase 10.1016/0039-6028(83)90095-X}
  {\bibfield  {journal} {\bibinfo  {journal} {Surface Science}\ }\textbf
  {\bibinfo {volume} {129}},\ \bibinfo {pages} {79} (\bibinfo {year}
  {1983})}\BibitemShut {NoStop}%
\bibitem [{\citenamefont {Ast}\ \emph {et~al.}(2016)\citenamefont {Ast},
  \citenamefont {J{\"a}ck}, \citenamefont {Senkpiel}, \citenamefont {Eltschka},
  \citenamefont {Etzkorn}, \citenamefont {Ankerhold},\ and\ \citenamefont
  {Kern}}]{ast_sensing_2016}%
  \BibitemOpen
  \bibfield  {author} {\bibinfo {author} {\bibfnamefont {C.~R.}\ \bibnamefont
  {Ast}}, \bibinfo {author} {\bibfnamefont {B.}~\bibnamefont {J{\"a}ck}},
  \bibinfo {author} {\bibfnamefont {J.}~\bibnamefont {Senkpiel}}, \bibinfo
  {author} {\bibfnamefont {M.}~\bibnamefont {Eltschka}}, \bibinfo {author}
  {\bibfnamefont {M.}~\bibnamefont {Etzkorn}}, \bibinfo {author} {\bibfnamefont
  {J.}~\bibnamefont {Ankerhold}}, \ and\ \bibinfo {author} {\bibfnamefont
  {K.}~\bibnamefont {Kern}},\ }\bibfield  {title} {\emph {\bibinfo {title}
  {Sensing the quantum limit in scanning tunnelling spectroscopy},}\ }\href
  {\doibase 10.1038/ncomms13009} {\bibfield  {journal} {\bibinfo  {journal}
  {Nature Communications}\ }\textbf {\bibinfo {volume} {7}},\ \bibinfo {pages}
  {13009} (\bibinfo {year} {2016})}\BibitemShut {NoStop}%
\bibitem [{\citenamefont {J\"ack}\ \emph {et~al.}(2016)\citenamefont {J\"ack},
  \citenamefont {Eltschka}, \citenamefont {Assig}, \citenamefont {Etzkorn},
  \citenamefont {Ast},\ and\ \citenamefont {Kern}}]{jack_critical_2016}%
  \BibitemOpen
  \bibfield  {author} {\bibinfo {author} {\bibfnamefont {B.}~\bibnamefont
  {J\"ack}}, \bibinfo {author} {\bibfnamefont {M.}~\bibnamefont {Eltschka}},
  \bibinfo {author} {\bibfnamefont {M.}~\bibnamefont {Assig}}, \bibinfo
  {author} {\bibfnamefont {M.}~\bibnamefont {Etzkorn}}, \bibinfo {author}
  {\bibfnamefont {C.~R.}\ \bibnamefont {Ast}}, \ and\ \bibinfo {author}
  {\bibfnamefont {K.}~\bibnamefont {Kern}},\ }\bibfield  {title} {\emph
  {\bibinfo {title} {Critical {Josephson} current in the dynamical {Coulomb}
  blockade regime},}\ }\href {\doibase 10.1103/PhysRevB.93.020504} {\bibfield
  {journal} {\bibinfo  {journal} {Phys. Rev. B}\ }\textbf {\bibinfo {volume}
  {93}},\ \bibinfo {pages} {020504} (\bibinfo {year} {2016})}\BibitemShut
  {NoStop}%
\bibitem [{\citenamefont {Falci}\ \emph {et~al.}(1991)\citenamefont {Falci},
  \citenamefont {Bubanja},\ and\ \citenamefont
  {Schön}}]{falci_quasiparticle_1991}%
  \BibitemOpen
  \bibfield  {author} {\bibinfo {author} {\bibfnamefont {G.}~\bibnamefont
  {Falci}}, \bibinfo {author} {\bibfnamefont {V.}~\bibnamefont {Bubanja}}, \
  and\ \bibinfo {author} {\bibfnamefont {G.}~\bibnamefont {Schön}},\
  }\bibfield  {title} {{\selectlanguage {english}\emph {\bibinfo {title}
  {Quasiparticle and {Cooper} pair tenneling in small capacitance {Josephson}
  junctions},}\ }}\href {\doibase 10.1007/BF01307643} {\bibfield  {journal}
  {\bibinfo  {journal} {Zeitschrift für Physik B Condensed Matter}\ }\textbf
  {\bibinfo {volume} {85}},\ \bibinfo {pages} {451} (\bibinfo {year}
  {1991})}\BibitemShut {NoStop}%
\bibitem [{\citenamefont {Roychowdhury}\ \emph {et~al.}(2015)\citenamefont
  {Roychowdhury}, \citenamefont {Dreyer}, \citenamefont {Anderson},
  \citenamefont {Lobb},\ and\ \citenamefont
  {Wellstood}}]{roychowdhury_microwave_2015}%
  \BibitemOpen
  \bibfield  {author} {\bibinfo {author} {\bibfnamefont {A.}~\bibnamefont
  {Roychowdhury}}, \bibinfo {author} {\bibfnamefont {M.}~\bibnamefont
  {Dreyer}}, \bibinfo {author} {\bibfnamefont {J.}~\bibnamefont {Anderson}},
  \bibinfo {author} {\bibfnamefont {C.}~\bibnamefont {Lobb}}, \ and\ \bibinfo
  {author} {\bibfnamefont {F.}~\bibnamefont {Wellstood}},\ }\bibfield  {title}
  {\emph {\bibinfo {title} {Microwave {Photon}-{Assisted} {Incoherent}
  {Cooper}-{Pair} {Tunneling} in a {Josephson} {STM}},}\ }\href {\doibase
  10.1103/PhysRevApplied.4.034011} {\bibfield  {journal} {\bibinfo  {journal}
  {Phys. Rev. Applied}\ }\textbf {\bibinfo {volume} {4}},\ \bibinfo {pages}
  {034011} (\bibinfo {year} {2015})}\BibitemShut {NoStop}%
\bibitem [{\citenamefont {Peters}\ \emph {et~al.}(2020)\citenamefont {Peters},
  \citenamefont {Bogdanoff}, \citenamefont {Acero~González}, \citenamefont
  {Melischek}, \citenamefont {Simon}, \citenamefont {Reecht}, \citenamefont
  {Winkelmann}, \citenamefont {von Oppen},\ and\ \citenamefont
  {Franke}}]{peters_resonant_2020}%
  \BibitemOpen
  \bibfield  {author} {\bibinfo {author} {\bibfnamefont {O.}~\bibnamefont
  {Peters}}, \bibinfo {author} {\bibfnamefont {N.}~\bibnamefont {Bogdanoff}},
  \bibinfo {author} {\bibfnamefont {S.}~\bibnamefont {Acero~González}},
  \bibinfo {author} {\bibfnamefont {L.}~\bibnamefont {Melischek}}, \bibinfo
  {author} {\bibfnamefont {J.~R.}\ \bibnamefont {Simon}}, \bibinfo {author}
  {\bibfnamefont {G.}~\bibnamefont {Reecht}}, \bibinfo {author} {\bibfnamefont
  {C.~B.}\ \bibnamefont {Winkelmann}}, \bibinfo {author} {\bibfnamefont
  {F.}~\bibnamefont {von Oppen}}, \ and\ \bibinfo {author} {\bibfnamefont
  {K.~J.}\ \bibnamefont {Franke}},\ }\bibfield  {title} {{\selectlanguage
  {english}\emph {\bibinfo {title} {Resonant {Andreev} reflections probed by
  photon-assisted tunnelling at the atomic scale},}\ }}\href {\doibase
  10.1038/s41567-020-0972-z} {\bibfield  {journal} {\bibinfo  {journal} {Nature
  Physics}\ }\textbf {\bibinfo {volume} {16}},\ \bibinfo {pages} {1222}
  (\bibinfo {year} {2020})}\BibitemShut {NoStop}%
\bibitem [{\citenamefont {Tien}\ and\ \citenamefont
  {Gordon}(1963)}]{tien_multiphoton_1963}%
  \BibitemOpen
  \bibfield  {author} {\bibinfo {author} {\bibfnamefont {P.~K.}\ \bibnamefont
  {Tien}}\ and\ \bibinfo {author} {\bibfnamefont {J.~P.}\ \bibnamefont
  {Gordon}},\ }\bibfield  {title} {\emph {\bibinfo {title} {Multiphoton
  {Process} {Observed} in the {Interaction} of {Microwave} {Fields} with the
  {Tunneling} between {Superconductor} {Films}},}\ }\href {\doibase
  10.1103/PhysRev.129.647} {\bibfield  {journal} {\bibinfo  {journal} {Physical
  Review}\ }\textbf {\bibinfo {volume} {129}},\ \bibinfo {pages} {647}
  (\bibinfo {year} {1963})}\BibitemShut {NoStop}%
\bibitem [{\citenamefont {Bae}\ \emph {et~al.}(2018)\citenamefont {Bae},
  \citenamefont {Yang}, \citenamefont {Willke}, \citenamefont {Choi},
  \citenamefont {Heinrich},\ and\ \citenamefont {Lutz}}]{bae_enhanced_2018}%
  \BibitemOpen
  \bibfield  {author} {\bibinfo {author} {\bibfnamefont {Y.}~\bibnamefont
  {Bae}}, \bibinfo {author} {\bibfnamefont {K.}~\bibnamefont {Yang}}, \bibinfo
  {author} {\bibfnamefont {P.}~\bibnamefont {Willke}}, \bibinfo {author}
  {\bibfnamefont {T.}~\bibnamefont {Choi}}, \bibinfo {author} {\bibfnamefont
  {A.~J.}\ \bibnamefont {Heinrich}}, \ and\ \bibinfo {author} {\bibfnamefont
  {C.~P.}\ \bibnamefont {Lutz}},\ }\bibfield  {title} {{\selectlanguage
  {english}\emph {\bibinfo {title} {Enhanced quantum coherence in exchange
  coupled spins via singlet-triplet transitions},}\ }}\href {\doibase
  10.1126/sciadv.aau4159} {\bibfield  {journal} {\bibinfo  {journal} {Science
  Advances}\ }\textbf {\bibinfo {volume} {4}},\ \bibinfo {pages} {eaau4159}
  (\bibinfo {year} {2018})}\BibitemShut {NoStop}%
\bibitem [{\citenamefont {Yang}\ \emph {et~al.}(2017)\citenamefont {Yang},
  \citenamefont {Bae}, \citenamefont {Paul}, \citenamefont {Natterer},
  \citenamefont {Willke}, \citenamefont {Lado}, \citenamefont {Ferrón},
  \citenamefont {Choi}, \citenamefont {Fernández-Rossier}, \citenamefont
  {Heinrich},\ and\ \citenamefont {Lutz}}]{yang_engineering_2017}%
  \BibitemOpen
  \bibfield  {author} {\bibinfo {author} {\bibfnamefont {K.}~\bibnamefont
  {Yang}}, \bibinfo {author} {\bibfnamefont {Y.}~\bibnamefont {Bae}}, \bibinfo
  {author} {\bibfnamefont {W.}~\bibnamefont {Paul}}, \bibinfo {author}
  {\bibfnamefont {F.~D.}\ \bibnamefont {Natterer}}, \bibinfo {author}
  {\bibfnamefont {P.}~\bibnamefont {Willke}}, \bibinfo {author} {\bibfnamefont
  {J.~L.}\ \bibnamefont {Lado}}, \bibinfo {author} {\bibfnamefont
  {A.}~\bibnamefont {Ferrón}}, \bibinfo {author} {\bibfnamefont
  {T.}~\bibnamefont {Choi}}, \bibinfo {author} {\bibfnamefont {J.}~\bibnamefont
  {Fernández-Rossier}}, \bibinfo {author} {\bibfnamefont {A.~J.}\ \bibnamefont
  {Heinrich}}, \ and\ \bibinfo {author} {\bibfnamefont {C.~P.}\ \bibnamefont
  {Lutz}},\ }\bibfield  {title} {\emph {\bibinfo {title} {Engineering the
  {Eigenstates} of {Coupled} {Spin}-1/2 {Atoms} on a {Surface}},}\ }\href
  {\doibase 10.1103/PhysRevLett.119.227206} {\bibfield  {journal} {\bibinfo
  {journal} {Physical Review Letters}\ }\textbf {\bibinfo {volume} {119}},\
  \bibinfo {pages} {227206} (\bibinfo {year} {2017})}\BibitemShut {NoStop}%
\bibitem [{\citenamefont {Steinbrecher}\ \emph {et~al.}(2021)\citenamefont
  {Steinbrecher}, \citenamefont {van Weerdenburg}, \citenamefont {Walraven},
  \citenamefont {van Mullekom}, \citenamefont {Gerritsen}, \citenamefont
  {Natterer}, \citenamefont {Badrtdinov}, \citenamefont {Rudenko},
  \citenamefont {Mazurenko}, \citenamefont {Katsnelson}, \citenamefont {van~der
  Avoird}, \citenamefont {Groenenboom},\ and\ \citenamefont
  {Khajetoorians}}]{steinbrecher_quantifying_2021}%
  \BibitemOpen
  \bibfield  {author} {\bibinfo {author} {\bibfnamefont {M.}~\bibnamefont
  {Steinbrecher}}, \bibinfo {author} {\bibfnamefont {W.~M.~J.}\ \bibnamefont
  {van Weerdenburg}}, \bibinfo {author} {\bibfnamefont {E.~F.}\ \bibnamefont
  {Walraven}}, \bibinfo {author} {\bibfnamefont {N.~P.~E.}\ \bibnamefont {van
  Mullekom}}, \bibinfo {author} {\bibfnamefont {J.~W.}\ \bibnamefont
  {Gerritsen}}, \bibinfo {author} {\bibfnamefont {F.~D.}\ \bibnamefont
  {Natterer}}, \bibinfo {author} {\bibfnamefont {D.~I.}\ \bibnamefont
  {Badrtdinov}}, \bibinfo {author} {\bibfnamefont {A.~N.}\ \bibnamefont
  {Rudenko}}, \bibinfo {author} {\bibfnamefont {V.~V.}\ \bibnamefont
  {Mazurenko}}, \bibinfo {author} {\bibfnamefont {M.~I.}\ \bibnamefont
  {Katsnelson}}, \bibinfo {author} {\bibfnamefont {A.}~\bibnamefont {van~der
  Avoird}}, \bibinfo {author} {\bibfnamefont {G.~C.}\ \bibnamefont
  {Groenenboom}}, \ and\ \bibinfo {author} {\bibfnamefont {A.~A.}\ \bibnamefont
  {Khajetoorians}},\ }\bibfield  {title} {\emph {\bibinfo {title} {Quantifying
  the interplay between fine structure and geometry of an individual molecule
  on a surface},}\ }\href {\doibase 10.1103/PhysRevB.103.155405} {\bibfield
  {journal} {\bibinfo  {journal} {Physical Review B}\ }\textbf {\bibinfo
  {volume} {103}},\ \bibinfo {pages} {155405} (\bibinfo {year}
  {2021})}\BibitemShut {NoStop}%
\bibitem [{\citenamefont {Seifert}\ \emph
  {et~al.}(2020{\natexlab{b}})\citenamefont {Seifert}, \citenamefont {Kovarik},
  \citenamefont {Juraschek}, \citenamefont {Spaldin}, \citenamefont
  {Gambardella},\ and\ \citenamefont {Stepanow}}]{seifert_longitudinal_2020}%
  \BibitemOpen
  \bibfield  {author} {\bibinfo {author} {\bibfnamefont {T.~S.}\ \bibnamefont
  {Seifert}}, \bibinfo {author} {\bibfnamefont {S.}~\bibnamefont {Kovarik}},
  \bibinfo {author} {\bibfnamefont {D.~M.}\ \bibnamefont {Juraschek}}, \bibinfo
  {author} {\bibfnamefont {N.~A.}\ \bibnamefont {Spaldin}}, \bibinfo {author}
  {\bibfnamefont {P.}~\bibnamefont {Gambardella}}, \ and\ \bibinfo {author}
  {\bibfnamefont {S.}~\bibnamefont {Stepanow}},\ }\bibfield  {title}
  {{\selectlanguage {english}\emph {\bibinfo {title} {Longitudinal and
  transverse electron paramagnetic resonance in a scanning tunneling
  microscope},}\ }}\href {\doibase 10.1126/sciadv.abc5511} {\bibfield
  {journal} {\bibinfo  {journal} {Science Advances}\ }\textbf {\bibinfo
  {volume} {6}},\ \bibinfo {pages} {eabc5511} (\bibinfo {year}
  {2020}{\natexlab{b}})}\BibitemShut {NoStop}%
\end{thebibliography}
\end{document}